# Inconsistencies in Simple Thermal Model Results for Near-Earth Asteroids between Infrared Telescope Facility SpeX and NEOWISE Data

Samuel A. Myers[1], Ellen S. Howell[1], Christopher Magri[2], Ronald J. Vervack, Jr.[3], Yanga R. Fernández[4], Mary L. Hinkle[4], and Sean E. Marshall[4]
[1] Lunar and Planetary Laboratory, University of Arizona, Tucson, AZ 85721, USA
[2] University of Maine Farmington, Farmington, ME 04938, USA
[3] Johns Hopkins Applied Physics Laboratory, Laurel, MD 20723, USA
[4] University of Central Florida, Orlando, FL 32816, USA

## Abstract

Understanding the properties of near-Earth asteroids (NEAs) is key for many aspects of planetary science, particularly planetary defense. Our current knowledge of NEA sizes and regolith properties is heavily dependent on simple thermal models. These models are often used to analyze data from missions such as NEOWISE because they are well suited to deal with large volumes of data. However, simple model results based on NEOWISE data may be inconsistent with results based on other types of observation in some cases. In this work, we seek to better understand these potential inconsistencies, as well as the situations for which they are most prevalent. We do this by comparing simple model results based on Infrared Telescope Facility SpeX data to similar results based on NEOWISE data. This is carried out for six NEAs that represent a range of spectral types, shapes, and rotation states. We find that models based on SpeX and NEOWISE data for these six objects are inconsistent in some cases, even though the SpeX results are consistent with other methods and observations. We find that objects observed at fainter magnitudes and objects with more primitive compositions are more likely to produce inconsistent fits. These results highlight the importance of better understanding the limitations of simple models as applied to large survey data sets like NEOWISE. This is particularly true as we move into an era where our understanding of the NEA population will be dominated by future large surveys such as NEO Surveyor.

*Unified Astronomy Thesaurus concepts:* Asteroids (72); Astronomical models (86); Near-Earth objects (1092)

## 1. Introduction

Studies of asteroids are important for answering some of the biggest questions in planetary science (e.g., A. Cellino et al. 2002; R. P. Binzel et al. 2015; M. Azadmanesh et al. 2023 and references therein). Knowledge of asteroids' sizes, albedo distributions, and regolith properties is key for our understanding of the solar system's formation, volatile delivery to the early Earth, small-body orbital evolution, the surfaces of airless bodies, linking meteorites to astronomical objects, and, of course, protecting Earth from potential threats posed by asteroids (e.g., A. W. Harris et al. 2015; P. Michel et al. 2015; M. Azadmanesh et al. 2023 and references therein). Studies of near-Earth asteroids (NEAs) are particularly important for these questions because their proximity to Earth makes them easier to observe. Therefore, it is critical that the methods we use to study these objects are robust. This includes both the types of data we collect and the models we use to understand them.

Historically, there have been numerous methods employed to investigate NEAs. One of the most extensively utilized has been thermal spectra, analyzed with various types of models (e.g., M. Delbo et al. 2015 and references therein). These models include complex thermophysical models, capable of incorporating full shape information about an object (e.g., C. Magri & E. S. Howell et al. 2018), and more simplified models. The two most prominent of these simple thermal models are the Standard Thermal Model (L. A. Lebofsky et al. 1986; L. A. Lebofsky & J. R. Spencer 1989) and the Near-Earth Asteroid Thermal Model (NEATM; A. W. Harris 1998). These models can be advantageous, as they run quickly and require only basic input parameters.

In order to operate so rapidly, however, these models incorporate simplifying assumptions about an asteroid's properties, such as its shape or surface properties, that can cause model results to be inaccurate or poorly constrained. Such assumptions are particularly relevant for measurements of asteroid sizes, as size is often estimated from a modeled albedo and absolute magnitude measurement (e.g., E. Bowell et al. 1989; M. Juric et al. 2002; P. Veres et al. 2015). Thus, size uncertainties can be influenced by uncertainties in the inferred albedos. As a result, NEAs that have only been analyzed with simple models are potentially prone to mischaracterization.

This is a particularly relevant concern as we move into the future of large asteroid surveys. One of the most anticipated future surveys is the NEO Surveyor mission. NEO Surveyor will be based on the same general architecture as NEOWISE but specifically tailored for NEA discovery. As such, it is anticipated to quickly become responsible for discovering a large percentage of the currently undiscovered population of NEAs with diameters between 140 m and 1 km (T. Grav et al. 2023; A. K. Mainzer et al. 2023). As a result, our future understanding of the NEA population will be heavily tied to the data taken by NEO Surveyor and its predecessor NEOWISE and the types of simple models used to understand those data. While it has been shown that these types of surveys constrain population properties well, current work has already shown that there can be issues when using these types of data and models to investigate the properties of individual objects. For example, studies have found inconsistencies in parameters derived with simple thermal models using NEOWISE data and those derived using other methods or other models (E. S. Howell & M. Nolan et al. 2012; P. Taylor et al. 2014; J. R. Masiero et al. 2019; P. Taylor et al. 2019; J. R. Masiero et al. 2021; T. G. Müller et al. 2023; S. A. Myers et al. 2023; J. Orlowski-Scherer et al. 2024). Better understanding the nature and sources of these inconsistencies will therefore be key for our ability to fully utilize the rich data sets we will soon have available and will allow us to understand NEAs both as a population and as individual objects.

In this paper, we set out to examine these inconsistencies, particularly when comparing simple thermal model results from ground-based thermal spectra and simple thermal model results based on photometry from NEOWISE. In particular, we aim to catalog the object characteristics that may make an object more prone to inconsistent results. To do this, we model spectral data of six different NEAs: (53319) 1999 JM8, (85989) 1999 JD6, (137032) 1998 UO1, (163899) 2003 SD220, (175706) 1996 FG3, and (285263) 1998 QE2. Hereafter, we refer to each asteroid by its nonyear provisional designation, e.g., JM8. We note that the analysis used here for QE2 was previously presented in S. A. Myers et al. (2023). These NEAs are observed from both the ground with the NASA Infrared Telescope Facility (IRTF) using the SpeX instrument and from space with the Wide-field Infrared Survey Explorer (WISE) and subsequent NEOWISE missions. (For convenience, we will refer to these data in general as NEOWISE data; however, it should be understood that four-band WISE data were analyzed as well, when available. The distinction between WISE and NEOWISE data will be called out when relevant.) All data are then modeled using a simple thermal model, and results are compared and analyzed. In this way we seek to identify and understand inconsistencies between model results based on the two types of data.

Section 2 describes the NEA data used in this work. We describe the simple thermal model we use in Section 3. In Section 4 a brief summary of each object is given, and the primary modeling results are presented. We discuss the primary results, including identified model fit inconsistencies, in Section 5. Finally, our key results are summarized and future work is laid out in Section 6.

## 2. Thermal Data

### 2.1. IRTF Data

All ground-based data were obtained using the SpeX instrument on the NASA IRTF in Hawai'i (J. T. Rayner et al. 2003). These observations were done as part of an ongoing investigation into the thermal and physical properties of NEAs. For all objects, we use spectral data normalized to 1.0 at ∼1.6 $\mu$m, where there is little thermal contribution in the spectrum. This approach allows us to take advantage of the lower relative uncertainties of normalized spectral data compared to absolute photometry. Each object is observed over many hours over multiple nights, allowing us to see the object under many different viewing geometries and illumination states. This ensures that we are able to infer the object's overall properties, not just properties from a single observation associated with a single surface view or rotational longitude. Fitting these observations therefore allows us to check for consistency in the fit parameters across a large portion of the asteroid's surface, ensuring that we are getting an accurate sense of the object's overall physical properties for comparison to other data.

We note that data were collected both before and after the SpeX upgrade in 2014 that increased the wavelength ranges in all modes. Data were collected in both the prism and long-wavelength cross-dispersed (LXD) modes for all objects. The prism wavelength ranges pre- and post-upgrade were 0.8–2.5 $\mu$m and 0.7–2.52 $\mu$m, respectively. Pre-upgrade we observed using LXD 1.9 (1.95–4.2 $\mu$m), and post-upgrade we observed using LXD long (1.98–5.3 $\mu$m). We note that in all cases the prism and LXD data overlap. Furthermore, all our observations cover the range from reflected-dominated to thermal-dominated light. This coverage ensures that our model is well constrained in both regimes. A slit width of 0.″8 and length of 15″ were used for all observations. We used exposure times of ∼10–30 s for all prism observations, ∼15 s for for pre-upgrade LXD observations, and 6 s (3 s, 2 co-adds) for post-upgrade LXD observations to avoid saturating the background at longer wavelengths.

Observations were done in pairs, nodding the telescope along the slit. In addition to the object, calibration stars were observed in a similar manner. These included a nearby G-type dwarf star with solar colors within ∼5° of the object and a well-characterized solar analog (C. D. Lewin et al. 2020). Details on the data processing procedure used can be found in E. Howell & C. Magri et al. (2018) and S. A. Myers et al. (2023), and a brief summary is given here. The Spextool software package (M. Cushing et al. 2004) was used to subtract the image pairs, apply flat-field corrections, sum images, and extract the spectra, which were then corrected for terrestrial atmospheric absorption lines. Notably, this atmospheric correction is done before any normalization, as we use a wavelength-dependent atmospheric model (S. D. Lord 1992). Bad data points were flagged throughout this process to be ignored in subsequent analysis. Asteroid spectra were then divided by the solar analog spectra to generate normalized spectral data. A weighted average was then used to combine each asteroid–star pair into a final spectrum that was binned down in wavelength to increase the signal-to-noise ratio. We use a relatively wide binning window because we are interested in the overall spectrum shape and do not require high spectral resolution. The prism spectra are binned by five channels for a final spectral resolution of 0.015 $\mu$m. The LXD spectra are binned by 25 channels for a final spectral resolution of 0.025 $\mu$m. Finally, the prism and LXD spectra from each night were joined together, and the resulting spectrum was normalized to 1.0 at ∼1.6 $\mu$m. This process was applied for all nights of observation.

Data from a single night were often broken up into several independent data sets, with each data set being assigned a letter and spanning a few minutes to a half hour of coverage. This approach allows us to look at different spectra as the objects rotate, giving us views of different parts of the surface. By fitting each of these spectra independently, we therefore ensure that we are getting a self-consistent view of the object's overall surface properties. Observing conditions for every SpeX data set are listed in Table 1, and all final, binned spectra are shown in Figure 1. Example SpeX prism spectra for each object are shown in Figure 2.

Table 1
Observing Conditions for All SpeX Observations

| Object | Date | Set | Midtime | $r_H$ (au) | $\Delta$ (au) | $\alpha$ (deg) | Upgrade? |
|---|---|---|---|---|---|---|---|
| JM8 | 2008 May 8 | A | 14:19:12 | 1.178 7 | 0.383 6 | 55.1 | Pre |
| JM8 | 2008 May 8 | B | 14:34:47 | 1.178 9 | 0.383 7 | 55.1 | Pre |
| JM8 | 2008 May 8 | C | 15:40:45 | 1.178 5 | 0.383 4 | 55.1 | Pre |
| JM8 | 2008 May 8 | D | 15:42:45 | 1.178 5 | 0.383 4 | 55.1 | Pre |
| JM8 | 2008 May 10 | A | 14:28:41 | 1.163 4 | 0.372 2 | 57.0 | Pre |
| JM8 | 2008 May 13 | A | 14:17:18 | 1.141 1 | 0.357 0 | 59.9 | Pre |
| JM8 | 2008 May 15 | A | 14:00:44 | 1.126 6 | 0.348 1 | 62.0 | Pre |
| JD6 | 2010 Jul 9 | A | 11:39:28 | 1.194 7 | 0.217 3 | 31.9 | Pre |
| JD6 | 2010 Jul 22 | A | 09:23:14 | 1.101 8 | 0.146 3 | 51.0 | Pre |
| JD6 | 2010 Jul 22 | B | 09:35:39 | 1.101 7 | 0.146 3 | 51.0 | Pre |
| JD6 | 2010 Jul 29 | A | 06:45:10 | 1.039 9 | 0.139 8 | 76.0 | Pre |
| JD6 | 2010 Jul 29 | B | 06:50:07 | 1.039 9 | 0.139 8 | 76.1 | Pre |
| JD6 | 2015 Jul 19 | A | 11:24:06 | 1.051 8 | 0.082 6 | 62.5 | Post |
| JD6 | 2015 Jul 19 | B | 11:55:40 | 1.051 6 | 0.082 4 | 62.6 | Post |
| JD6 | 2015 Jul 21 | A | 11:14:52 | 1.033 0 | 0.065 1 | 73.2 | Post |
| UO1 | 2008 Oct 1 | A | 07:14:42 | 1.050 5 | 0.097 3 | 57.1 | Pre |
| UO1 | 2008 Oct 3 | A | 05:23:57 | 1.077 5 | 0.122 7 | 48.6 | Pre |
| UO1 | 2008 Oct 3 | B | 05:58:17 | 1.077 9 | 0.123 1 | 48.5 | Pre |
| UO1 | 2008 Oct 3 | C | 06:18:50 | 1.078 0 | 0.123 2 | 48.4 | Pre |
| UO1 | 2008 Oct 3 | D | 06:26:31 | 1.078 1 | 0.123 3 | 48.4 | Pre |
| UO1 | 2008 Oct 3 | E | 06:37:21 | 1.078 2 | 0.123 4 | 48.4 | Pre |
| UO1 | 2008 Oct 3 | F | 06:42:31 | 1.078 3 | 0.123 5 | 48.4 | Pre |
| UO1 | 2010 Oct 3 | A | 05:46:27 | 0.996 3 | 0.090 3 | 90.2 | Pre |
| UO1 | 2010 Oct 3 | B | 06:14:04 | 0.996 6 | 0.090 4 | 90.0 | Pre |
| UO1 | 2010 Oct 3 | C | 06:46:01 | 0.996 9 | 0.090 6 | 89.8 | Pre |
| UO1 | 2010 Oct 8 | A | 05:27:13 | 1.067 3 | 0.149 0 | 59.2 | Pre |
| UO1 | 2010 Oct 8 | B | 05:53:02 | 1.067 5 | 0.149 2 | 59.2 | Pre |
| UO1 | 2010 Oct 8 | C | 06:21:28 | 1.067 8 | 0.149 5 | 59.1 | Pre |
| UO1 | 2010 Oct 14 | A | 06:19:37 | 1.150 0 | 0.244 2 | 46.4 | Pre |
| UO1 | 2010 Oct 14 | B | 07:17:43 | 1.150 5 | 0.244 9 | 46.4 | Pre |
| SD220 | 2018 Nov 20 | A | 15:46:52 | 1.001 7 | 0.108 1 | 79.6 | Post |
| SD220 | 2018 Nov 28 | A | 15:26:36 | 0.998 8 | 0.083 3 | 79.1 | Post |
| SD220 | 2018 Dec 4 | A | 15:20:59 | 0.994 1 | 0.063 9 | 80.5 | Post |
| SD220 | 2018 Dec 7 | A | 15:33:08 | 0.990 9 | 0.054 2 | 82.3 | Post |
| SD220 | 2021 Dec 16 | A | 05:14:08 | 0.982 0 | 0.036 5 | 92.3 | Post |
| SD220 | 2021 Dec 16 | B | 05:47:45 | 0.981 9 | 0.036 5 | 92.4 | Post |
| SD220 | 2021 Dec 24 | A | 05:50:05 | 0.967 9 | 0.041 7 | 110.9 | Post |
| SD220 | 2021 Dec 24 | B | 06:04:41 | 0.967 8 | 0.041 7 | 110.9 | Post |
| FG3 | 2011 Nov 28 | A | 15:21:15 | 1.032 0 | 0.103 7 | 61.4 | Pre |
| FG3 | 2011 Dec 6 | C | 13:08:28 | 1.078 3 | 0.117 8 | 36.0 | Pre |
| FG3 | 2011 Dec 6 | A | 13:23:02 | 1.078 3 | 0.117 9 | 35.9 | Pre |
| FG3 | 2011 Dec 6 | D | 13:39:51 | 1.078 4 | 0.117 9 | 35.9 | Pre |
| FG3 | 2011 Dec 6 | E | 14:44:45 | 1.078 7 | 0.118 0 | 35.8 | Pre |
| FG3 | 2011 Dec 6 | B | 14:59:44 | 1.078 7 | 0.118 0 | 35.7 | Pre |
| FG3 | 2011 Dec 6 | F | 15:10:07 | 1.078 8 | 0.118 1 | 35.7 | Pre |
| FG3 | 2011 Dec 24 | B | 10:38:49 | 1.174 2 | 0.192 4 | 7.2 | Pre |
| FG3 | 2011 Dec 24 | A | 10:47:42 | 1.174 2 | 0.192 4 | 7.2 | Pre |
| FG3 | 2011 Dec 24 | C | 10:56:35 | 1.174 3 | 0.192 4 | 7.2 | Pre |
| FG3 | 2012 Jan 4 | A | 07:37:45 | 1.225 6 | 0.260 5 | 19.2 | Pre |
| FG3 | 2022 Apr 11 | A | 09:46:17 | 1.142 2 | 0.145 1 | 14.0 | Post |
| FG3 | 2022 Apr 11 | B | 10:08:53 | 1.142 1 | 0.144 9 | 14.0 | Post |
| FG3 | 2022 Apr 11 | C | 10:26:25 | 1.142 0 | 0.144 9 | 14.0 | Post |
| FG3 | 2022 Apr 19 | A | 08:52:40 | 1.099 0 | 0.109 0 | 28.2 | Post |
| FG3 | 2022 Apr 19 | B | 09:07:03 | 1.099 0 | 0.109 0 | 28.3 | Post |
| FG3 | 2022 Apr 19 | C | 09:26:54 | 1.098 9 | 0.108 9 | 28.3 | Post |
| FG3 | 2022 Apr 28 | C | 07:47:08 | 1.047 6 | 0.083 2 | 58.6 | Post |
| QE2 | 2013 May 3 | A | 06:46:50 | 1.046 8 | 0.040 3 | 34.3 | Pre |
| QE2 | 2013 May 3 | B | 07:22:08 | 1.046 8 | 0.040 3 | 34.2 | Pre |
| QE2 | 2013 May 3 | C | 08:36:57 | 1.049 8 | 0.040 2 | 33.9 | Pre |
| QE2 | 2013 Jun 2 | A | 06:51:57 | 1.052 2 | 0.040 1 | 18.3 | Pre |

**Table 1**
(Continued)

| Object | Date | Set | Midtime | $r_H$ (au) | $\Delta$ (au) | $\alpha$ (deg) | Upgrade? |
|---|---|---|---|---|---|---|---|
| QE2 | 2013 Jun 2  | B | 07:08:19 | 1.052 2 | 0.040 1 | 18.3 | Pre |
| QE2 | 2013 Jun 2  | C | 07:17:50 | 1.052 2 | 0.040 1 | 18.3 | Pre |
| QE2 | 2013 Jun 2  | D | 07:34:17 | 1.052 2 | 0.040 1 | 18.2 | Pre |
| QE2 | 2013 Jun 8  | A | 08:12:16 | 1.067 1 | 0.060 5 | 30.0 | Pre |
| QE2 | 2013 Jun 8  | B | 09:25:01 | 1.067 2 | 0.060 8 | 30.1 | Pre |
| QE2 | 2013 Jun 8  | C | 09:37:14 | 1.067 2 | 0.060 8 | 30.1 | Pre |
| QE2 | 2013 Jun 8  | D | 10:38:10 | 1.067 4 | 0.061 0 | 30.2 | Pre |
| QE2 | 2013 Jun 8  | E | 10:50:40 | 1.067 4 | 0.061 1 | 30.2 | Pre |
| QE2 | 2013 Jun 15 | A | 11:06:28 | 1.091 0 | 0.098 8 | 38.8 | Pre |
| QE2 | 2013 Jun 15 | B | 12:16:11 | 1.091 2 | 0.099 1 | 38.8 | Pre |
| QE2 | 2013 Jun 18 | A | 13:07:51 | 1.103 3 | 0.116 9 | 39.7 | Pre |
| QE2 | 2013 Jul 10 | A | 10:23:08 | 1.218 8 | 0.256 2 | 34.0 | Pre |
| QE2 | 2013 Jul 10 | B | 10:29:19 | 1.218 9 | 0.256 2 | 34.0 | Pre |
| QE2 | 2013 Jul 10 | C | 11:10:20 | 1.219 0 | 0.256 4 | 34.0 | Pre |
| QE2 | 2013 Jul 10 | D | 11:49:53 | 1.219 2 | 0.256 6 | 34.0 | Pre |
| QE2 | 2013 Jul 10 | E | 13:09:29 | 1.219 6 | 0.257 0 | 34.0 | Pre |

**Note.** "Set" refers to the different data sets on a given night. For some FG3 dates the set lettering is not in alphabetical order across time, due to the way in which the data were processed. We note that on 2022 April 28 for FG3 set C was the only viable data set of the evening, due to weather. Furthermore, we note that 2015 July 21 for JD6 was provided to us by V. Reddy (personal communication) and has post-upgrade LXD short (1.67–4.2 μm) data only. "Midtime" refers to the midtime of the observations in UTC. $r_H$ is the Sun–object distance, $\Delta$ is the Earth–object distance, and $\alpha$ is the solar phase angle. "Upgrade?" denotes whether the observations were done before (pre) or after (post) the upgrades to the SpeX instrument in 2014.

### 2.2. NEOWISE Data

For every object observed with SpeX, models were also compared to data taken by NEOWISE. All NEOWISE (A. Mainzer et al. 2011, 2014a) data were taken from the NASA/IPAC Infrared Science Archive and associated catalogs (NEOWISE Team 2020; WISE Team 2020a, 2020b). NEOWISE is a space-based platform that measures thermal flux, producing absolute photometry. NEOWISE started life as the WISE telescope, observing in four bands, W1–W4, with effective wavelengths of 3.4, 4.6, 12, and 22 μm, respectively. These observations were made simultaneously, splitting the light among the four separate detectors. Following depletion of the cryogenic coolant, the mission initially continued in its "post-cryo" phase, taking data only in the W1 and W2 bands. Together, the full WISE and post-cryo phases spanned roughly a year between 2010 January and 2011 February. The spacecraft was then reactivated as NEOWISE in 2013 December, continuing observations in both the W1 and W2 bands. We used all available data for each object. Thus, we used data from each of the periods of the mission life: WISE full four-band cryo (referred to here as WISE or WA), post-cryo with only the W1 and W2 bands (referred to here as post-cryo or PC), and the NEOWISE regular mission with the W1 and W2 bands (referred to here as NEOWISE or NW).

We work with data that were extracted and processed by the NEOWISE automatic pipeline rather than extracting from the raw images ourselves. Data were checked against the tracklets submitted to the Minor Planet Center (MPC) to ensure accurate data retrieval for moving objects. This was done by retrieving observations from the MPC initially and then searching the L1b Single Source Catalogs on the retrieved object positions with a 0.″3 radius and restricting to observations that match within 2 s of the reported observation time from the MPC. This was done following the process laid out in the WISE Data Processing Handbook (E. L. Wright et al. 2010). We then removed data points that were flagged for contamination or overexposure and data points that had null detections in any band. Data were then converted from NEOWISE magnitudes to $F_\lambda$ units following the procedures outlined in the WISE Data Processing Handbook.

The WISE observations require a color correction that depends on the input spectral flux distribution. For stellar sources, a blackbody temperature within the filter band is listed in the WISE Data Processing Handbook, along with appropriate correction factors. However, asteroids instead have a distribution of temperatures on their viewed surfaces, and thus their spectra are a sum of many different blackbody spectra. Particularly for W1, this can lead to a wide range of possible correction factors, making the selection of a color correction for a given data set nontrivial.

For each data set, we calculate the color correction factors according to the following process. First, an initial blackbody temperature was calculated using the blackbody temperature relation for an isothermal object, assuming an emissivity of unity,

$$\sigma_{\rm sb} T^4 = \frac{L_\odot (1 - A)}{16\pi r_H^2}, \quad (1)$$

where $\sigma_{\rm sb}$ is the Stefan–Boltzmann constant, $L_\odot$ is the solar luminosity, $A$ is the Bond albedo, and $r_H$ is the Sun–object distance. The Bond albedo is estimated using the method described in L. A. Lebofsky & J. R. Spencer (1989):

$$A = (0.29 + 0.684G)p. \quad (2)$$

Here $p$ is the visual geometric albedo, taken from the model fits to the SpeX data for each object, and $G$ is the slope parameter in the $HG$ magnitude system (E. Bowell et al. 1989). Reported $G$ values are used when available; otherwise, an assumption of $G = 0.15$ is used. (We note that the final correction temperature is only sensitive to changes in $G$ by at most ∼10 K, less than other uncertainties in the color correction temperature (see

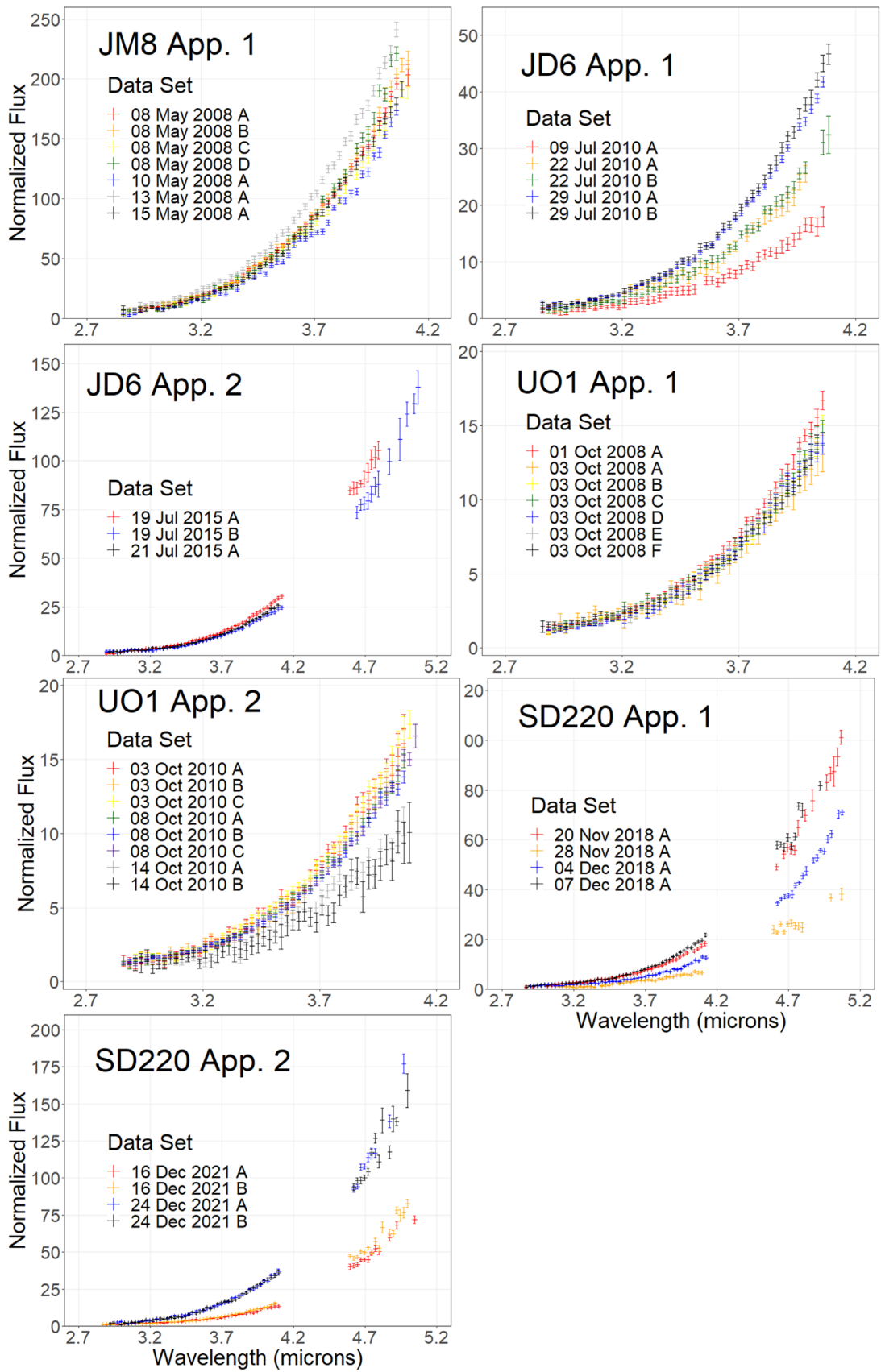

**Figure 1.** Final, binned SpeX spectra. Note that the portion of each spectrum below 2.7 $\mu$m is omitted for clarity. Example SpeX prism spectra for each object are shown in Figure 2. Each panel shows all data for a different apparition of each object. Apparition 1 for both FG3 and QE2 has been further separated for clarity. We note that on 2022 April 28 for FG3 set C was the only viable data set of the evening, due to weather, and that data were not available at wavelengths longer than 4.2 $\mu$m. Furthermore, we note that 2015 July 21 for JD6 Apparition 2 was provided to us by V. Reddy (personal communication) and has LXD short (1.67–4.2 $\mu$m) data only. The y-axis is normalized flux, normalized to 1.0 at ~1.6 $\mu$m. The x-axis is wavelength in microns. Note that different panels have different axis scales.

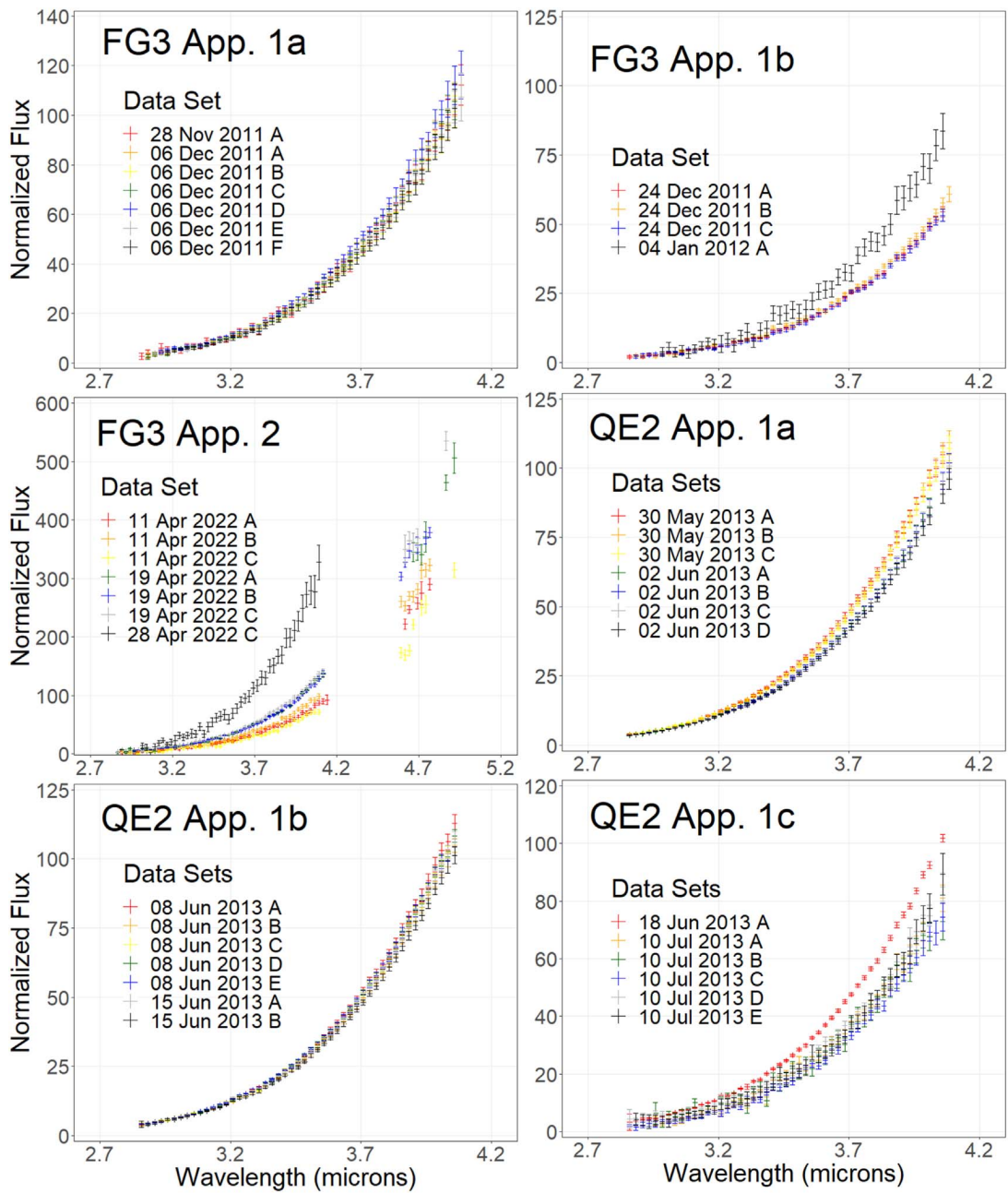

**Figure 1.** (Continued.)

below). This initial temperature used for each data set is given in Table 2.

However, the objects we are looking at are not isothermal, but are instead a sum of a range of temperatures over the viewed area. To account for this, we examined how a blackbody curve calculated using the temperature derived from the isothermal formula above compared to the best-fit effective blackbody curve that fit our SpeX spectra. This process was done for a range of viewing geometries that were representative of the geometries seen in our NEOWISE data. The objects tend to be near 1 au and at high solar phase angles, or at moderate solar phase angles at larger distances (see Table 2 and Figure 3). Furthermore, the isothermal assumptions cancel with corrections based on the phase angle. Thus, because the phase angle and solar distance are so highly correlated for the NEOWISE data, the following procedure is sufficient out to roughly 2 au, beyond which we have no NEOWISE data.

We found that the differences were not more than 30 K for the geometries we sampled, with the isothermal assumption always being less than the matched blackbody curve at greater than roughly 1.2 au. Therefore, for data at distances greater than 1.2 au we add the additional 30 K uncertainty to the isothermal assumption to reflect the uncertainty in the blackbody temperature. Thus, we effectively assume a 30 K uncertainty range in temperature, with the lower end given by the isothermal calculation. For data at distances less than 1.2 au the 30 K uncertainty interval is centered on the guess temperature. Because of the correlation between phase angle and solar distance discussed above, this allows us to account for the correct asteroid temperature, despite the inherent insufficiency of the isothermal assumption. This uncertainty range in temperature translates to an uncertainty range in the applied color correction, which in turn increases the uncertainty on the data. Finally, additional corrections were applied to the W4 point when applicable (E. L. Wright et al. 2010). Once these corrections are all applied, this process results in all flux densities being reported in units of W cm$^{-2}$ $\mu$m$^{-1}$. Additional details on parts of this process are described in S. A. Myers et al. (2023).

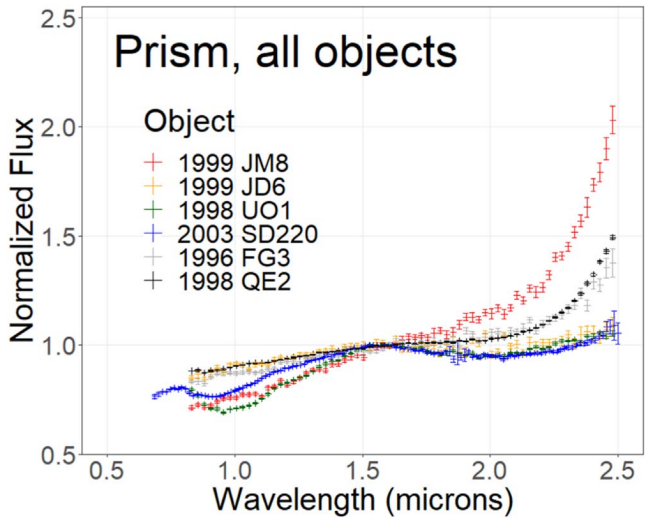

**Figure 2.** Example SpeX prism spectrum for each object. The *y*-axis is normalized flux. The *x*-axis is wavelength in microns. Data were taken on 2008 May 13 (JM8), 2010 July 29 (JD6), 2010 October 3 (UO1), 2018 December 7 (SD220), 2011 November 28 (FG3), and 2013 June 8 (QE2).

After conversion, all observations within a set were then averaged together, averaging each band independently of the others. The final data points are given by the weighted average of the observations in the band. The final 1$\sigma$ uncertainties are given by the variance of the observations in the given band, divided by the square root of the number of observations minus 1. This approach is appropriate, as NEOWISE data are dominated by systematic errors, such as light-curve variations at different (unknown) rotational phases.

All data taken across a roughly 24 hr period, or all data taken during a single apparition, whichever is shorter, were averaged together in this way. Nearly all data were taken at a cadence longer than the object's rotation period, so we did not attempt to average together data taken at similar rotation states. Natural gaps in the data were used as the break point between averaged data sets when possible. Final averaged data sets are identified by three letters. The first two letters denote the mission phase the data were taken from, and the last letter is a unique identifier. Final NEOWISE converted data are shown in Figure 4, and their associated observing conditions are listed in Table 2.

As a final check on the data quality, we visually inspect all raw images that contributed to our NEOWISE data. Images were inspected for null detections and possible contamination. This inspection revealed that many problematic images were not correctly flagged by the NEOWISE automatic data processing pipeline. Ultimately, we entirely rejected 9 out of our original 68 data sets, or about 13%, because all images were null detections or contaminated. (This is why some letters are missing in the data sets listed in Table 2 and Figure 4.) Roughly half of the remaining data sets contained at least one image that was potentially problematic. These data sets were adjusted by removing the problematic images from the final averaged spectrum. We note that overall the process laid out in the WISE Data Processing Handbook (E. L. Wright et al. 2010) was not able to identify a meaningful percentage of problematic images and that inspection of the raw images was needed to ensure that only real detections were being used in the analysis.

### 3. Thermal Model

#### 3.1. Model Description

We fit every SpeX and NEOWISE data set with a simple thermal model based on the Standard Thermal Model (L. A. Lebofsky et al. 1986; L. A. Lebofsky & J. R. Spencer 1989; J. R. Spencer et al. 1989) and NEATM (A. W. Harris 1998). We refer to this model as the NEATM-like model. (For more details see E. Howell & C. Magri et al. 2018 and S. A. Myers et al. 2023.) In general, the model functions similarly to NEATM in that it assumes a spherical shape for the object and equatorial viewing and illumination geometries. Unlike NEATM, however, the model includes a rudimentary incorporation of the thermal inertia of the object. The model assumes a fixed, prograde rotation rate for the object that allows the model to track differences in thermal emission across the dayside and nightside of the object. In this way, an estimate of the thermal inertia, or how well the object's surface is retaining heat energy from the Sun, is calculated. We note that many NEAs are known to have retrograde rotations; however, this incorporation still allows for heat transport across the surface, which allows for fitting a wider range of fluxes to determine the acceptable parameter ranges for good fits. Thus, even for retrograde objects, this incorporation still allows for a more realistic accounting of surface heat transport. Furthermore, combined with our observations taken across multiple rotation phases and viewing geometries for each object, this allows us to constrain the object's overall thermal properties well, not just its properties as they appear in one particular instance. We note that in cases where there is a secondary, the secondary is ignored. This is done because in all cases here the primary dominates the emissions. For example, FG3 is a binary, but the secondary only contributes ∼10% of the total flux.

Additionally, our NEATM-like model employs a different scheme than NEATM for handling phase angle corrections. The NEATM model uses the apparent illuminated area at the observed solar phase angle to estimate the phase angle correction for a smooth surface. The NEATM-like model uses a phase angle correction that is equivalent to a slightly rough surface 15% covered by hemispherical craters. These differences result in our NEATM-like model overestimating the object flux relative to NEATM at a given phase angle. We note that these differences are only relevant for our fits to the NEOWISE absolute flux data, as absolute magnitude effects are divided out of the normalized flux SpeX data. These differences are small ($\lesssim$50%) at phase angles of $\leqslant$50°. Variations in this range are within typical uncertainties for absolute flux data, particularly at unknown rotational phases with large light-curve amplitudes. At phase angles of >50° the effects can be larger, with differences between our model and NEATM of up to a factor of two. In this work, phase angles in this range are only relevant for NEOWISE observations of SD220 and UO1. However, we note that we find consistency

Table 2
Observing Conditions for All NEOWISE Observations

| Object | Set | Cntr U | Date | Midtime | $r_H$ (au) | $\Delta$ (au) | $\alpha$ (deg) | $T$ (K) | Consistent? |
|---|---|---|---|---|---|---|---|---|---|
| JM8 | NWC | 78–82 | 2017 Feb 18 | 06:54:27 | 1.698 8 | 1.416 5 | 35.5 | 213 | C |
| JM8 | NWD | 84–92 | 2017 Feb 26 | 12:56:31 | 1.627 1 | 1.258 3 | 37.5 | 218 | C |
| JM8 | NWE | 93–99, 101–104, 106–109 | 2017 Feb 27 | 16:55:01 | 1.616 9 | 1.267 4 | 37.8 | 219 | C |
| JM8 | NWF | 110–119 | 2017 Feb 28 | 22:27:50 | 1.606 2 | 1.248 7 | 38.1 | 219 | C |
| JM8 | NWG | 121, 123–125, 127–131 | 2017 Apr 12 | 02:49:56 | 1.250 3 | 0.754 6 | 53.3 | 248 | C |
| JM8 | NWH | 132–135, 137, 139–144 | 2017 Apr 13 | 06:28:05 | 1.241 3 | 0.746 1 | 53.9 | 249 | C |
| JM8 | NWI | 146–149, 152–155, 157, 158, 160, 161 | 2017 Apr 14 | 05:12:28 | 1.234 0 | 0.739 5 | 54.3 | 250 | C |
| JM8 | NWJ | 164–166, 168, 169 | 2017 Apr 15 | 10:41:57 | 1.224 6 | 0.731 0 | 55.0 | 251 | C |
| JM8 | NWK | 171–175, 177–185 | 2017 Aug 28 | 12:06:01 | 1.433 7 | 1.018 8 | 44.8 | 232 | C |
| JM8 | NWL | 187–190 | 2017 Aug 29 | 07:10:50 | 1.440 5 | 1.019 2 | 44.5 | 232 | C |
| JD6 | WAA | 5, 6 | 2010 Apr 25 | 04:41:46 | 1.440 5 | 1.039 3 | 44.3 | 228 | N |
| JD6 | WAB | 13, 14, 18, 20, 21 | 2010 Apr 25 | 22:47:39 | 1.440 7 | 1.031 4 | 44.3 | 228 | N |
| JD6 | WAD | 24–27 | 2010 Jul 29 | 18:57:19 | 1.035 1 | 0.140 6 | 78.0 | 269 | C |
| JD6 | NWA | 1 | 2014 Aug 4 | 01:49:46 | 1.354 9 | 0.806 4 | 48.2 | 235 | C |
| JD6 | NWB | 3, 4, 6, 8, 10, 11, 13, 14 | 2014 Aug 9 | 04:57:52 | 1.332 9 | 0.839 5 | 49.5 | 237 | C |
| JD6 | NWC | 21, 34 | 2015 Apr 30 | 12:23:01 | 1.436 6 | 1.031 5 | 44.5 | 228 | N |
| JD6 | NWE | 43, 47 | 2019 Apr 12 | 19:57:04 | 1.263 1 | 0.770 8 | 52.5 | 243 | C |
| JD6 | NWF | 56, 57, 60–64 | 2019 Jul 21 | 12:45:38 | 1.377 7 | 0.646 7 | 43.9 | 233 | C |
| JD6 | NWH | 92, 97, 99, 101–105 | 2023 May 30 | 09:05:33 | 1.218 6 | 0.476 7 | 53.9 | 248 | C |
| JD6 | NWI | 107, 108, 110–113 | 2023 May 30 | 22:43:13 | 1.222 4 | 0.481 7 | 52.6 | 247 | C |
| JD6 | NWJ | 120, 121, 127, 129, 131, 132, 134 | 2023 May 31 | 21:51:35 | 1.228 7 | 0.490 2 | 53.2 | 247 | C |
| JD6 | NWK | 137, 139, 140 | 2023 Jun 1 | 18:51:03 | 1.234 4 | 0.497 9 | 52.8 | 246 | C |
| JD6 | NWL | 143–149 | 2023 Jun 2 | 13:14:38 | 1.239 3 | 0.497 9 | 52.5 | 246 | C |
| UO1 | PCA | 2 | 2010 Oct 4 | 07:09:29 | 1.011 5 | 0.099 3 | 80.7 | 269 | C |
| UO1 | NWA | 16, 17, 19 | 2014 Oct 18 | 00:19:21 | 1.052 7 | 0.297 9 | 71.0 | 263 | I |
| UO1 | NWD | 76 | 2016 Nov 9 | 03:11:49 | 1.281 0 | 0.655 1 | 49.5 | 239 | C |
| SD220 | NWA | 2–4, 6–9 | 2015 Nov 17 | 00:55:04 | 1.001 9 | 0.165 1 | 80.7 | 270 | C |
| SD220 | NWB | 11, 12, 14–26 | 2015 Nov 18 | 01:14:52 | 1.001 8 | 0.162 0 | 80.7 | 270 | I |
| SD220 | NWC | 27–37, 39 | 2015 Nov 18 | 20:06:34 | 1.001 6 | 0.159 6 | 80.7 | 270 | N |
| SD220 | NWD | 41–50 | 2015 Nov 20 | 02:40:10 | 1.001 2 | 0.155 7 | 80.7 | 270 | C |
| SD220 | NWE | 52–56, 58–68 | 2018 Nov 10 | 22:23:42 | 0.999 7 | 0.136 2 | 82.1 | 270 | C |
| SD220 | NWF | 69–78 | 2018 Nov 11 | 18:40:16 | 1.000 1 | 0.133 8 | 81.8 | 270 | C |
| SD220 | NWG | 79, 80, 83 | 2018 Dec 11 | 17:28:48 | 0.985 7 | 0.041 2 | 87.3 | 272 | C |
| SD220 | NWH | 87, 88, 90, 91, 93, 94 | 2021 Nov 10 | 05:05:41 | 0.996 8 | 0.117 4 | 83.4 | 270 | C |
| SD220 | NWI | 96–101, 104, 105 | 2021 Nov 10 | 20:26:53 | 0.997 3 | 0.115 8 | 83.1 | 270 | C |
| FG3 | WAA | 1–4, 6–17 | 2010 Apr 30 | 16:14:37 | 1.219 8 | 0.611 4 | 55.4 | 252 | N |
| FG3 | PCA | 1, 3–6, 8, 11–15 | 2010 Nov 19 | 03:58:57 | 1.149 8 | 0.594 9 | 59.3 | 259 | I |
| FG3 | NWA | 2–4 | 2021 Feb 3 | 21:00:15 | 1.302 2 | 0.886 0 | 49.2 | 244 | I |
| FG3 | NWB | 12, 14, 15, 17, 19, 21 | 2021 Feb 13 | 20:33:40 | 1.264 7 | 0.787 1 | 51.3 | 247 | I |
| FG3 | NWC | 23, 25–27, 29–31, 33–35, 37, 38, 40, 42–48 | 2021 Feb 15 | 11:04:57 | 1.258 2 | 0.771 6 | 51.7 | 248 | I |
| FG3 | NWD | 51, 53, 56, 68 | 2021 Feb 17 | 03:00:45 | 1.251 3 | 0.755 7 | 52.1 | 248 | I |
| FG3 | NWE | 60–62, 64–67, 70, 72–75 | 2021 Mar 7 | 08:30:30 | 1.167 4 | 0.595 2 | 58.2 | 257 | I |
| FG3 | NWG | 82, 85, 90 | 2021 Mar 9 | 14:38:13 | 1.155 9 | 0.577 7 | 59.2 | 259 | N |
| FG3 | NWH | 95, 96, 107, 109 | 2021 Mar 11 | 09:15:36 | 1.146 7 | 0.564 3 | 60.0 | 260 | N |
| FG3 | NWI | 114, 115, 117 | 2021 Dec 31 | 22:17:14 | 1.421 8 | 1.017 0 | 43.8 | 233 | I |
| FG3 | NWJ | 123, 124, 128, 130 | 2022 Jan 1 | 20:22:14 | 1.421 6 | 1.008 1 | 43.8 | 233 | N |
| FG3 | NWK | 133, 134 | 2022 Apr 29 | 23:14:08 | 1.037 9 | 0.080 7 | 65.6 | 273 | C |
| FG3 | NWL | 136–139 | 2022 Dec 7 | 03:35:25 | 1.357 3 | 0.953 0 | 46.5 | 239 | I |
| FG3 | NWM | 144, 148, 152 | 2022 Dec 8 | 00:57:00 | 1.359 6 | 0.946 9 | 46.4 | 238 | N |
| FG3 | NWN | 157, 159–166 | 2023 Apr 14 | 11:49:59 | 1.267 9 | 0.488 6 | 47.3 | 247 | C |
| FG3 | NWO | 167–172 | 2023 Nov 21 | 08:04:08 | 1.185 4 | 0.661 9 | 56.4 | 255 | C |
| FG3 | NWP | 174–178, 180, 181 | 2023 Nov 21 | 23:28:01 | 1.188 6 | 0.659 9 | 56.2 | 255 | I |
| QE2 | NWA | 66, 68, 70, 79–81 | 2017 Jul 1 | 10:43:44 | 1.767 6 | 1.445 3 | 35.1 | 221 | I |

**Note.** "Set" refers to the different data sets for a given object. The first two letters denote whether the observation was made during the WISE (WA), post-cyro (PC), or NEOWISE (NW) phase of the mission. The third letter denotes the set. Note that missing letters indicate sets that were rejected after looking at the raw NEOWISE images (Section 2.2). Cntr U values are the frame numbers assigned to the individual NEOWISE observations that make up each data set (R. Cutri et al. 2015). "Midtime" refers to the midtime of the observations in UTC. $r_H$ is the Sun–object distance, $\Delta$ is the Earth–object distance, and $\alpha$ is the solar phase angle. $T$ is the blackbody correction temperature used in the data processing (Section 2.2). "Consistent?" denotes whether or not the NEOWISE modeling results are consistent with the SpeX modeling results, with "C" denoting consistent, "I" denoting inconsistent, and "N" denoting no fit (Section 5).

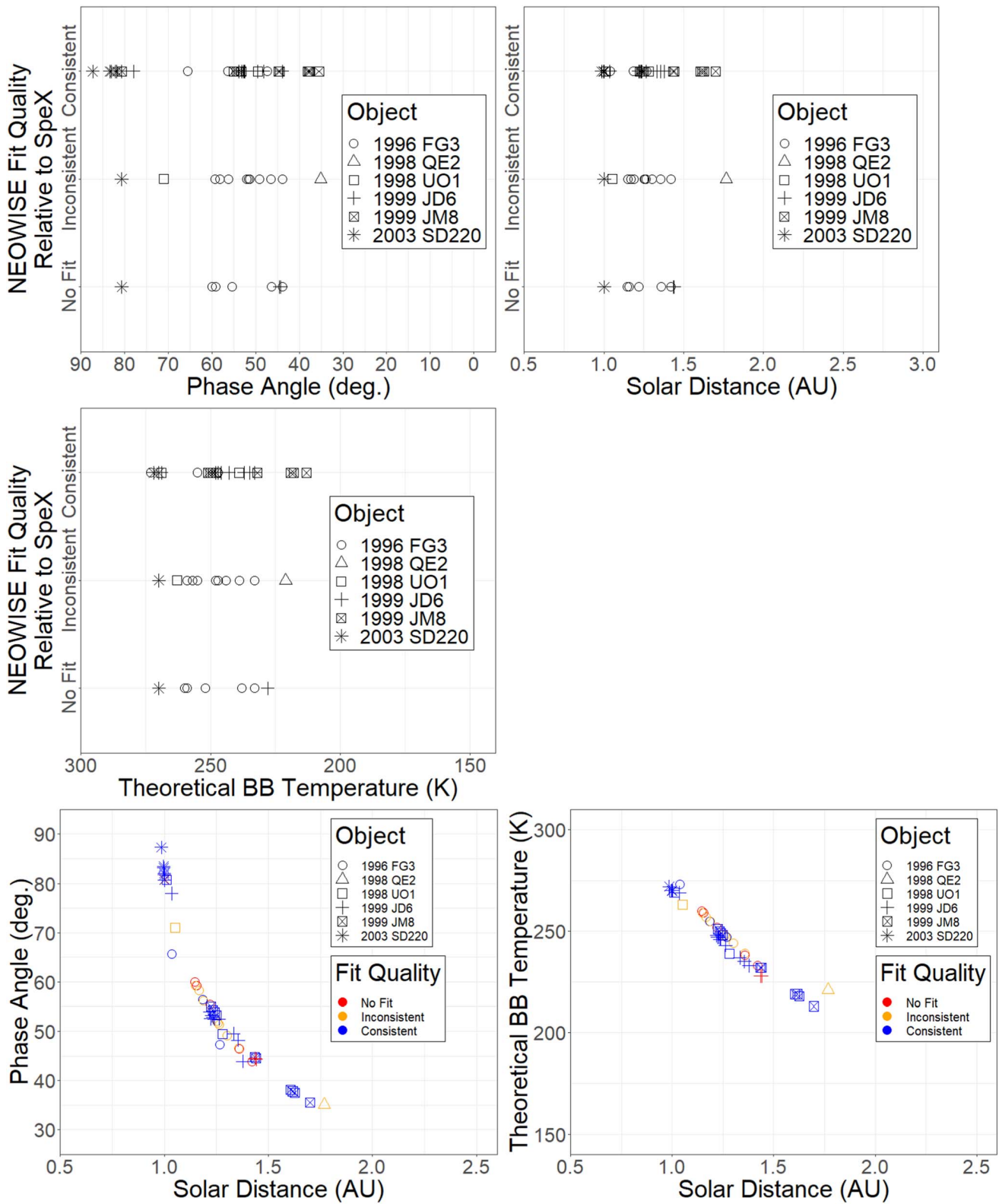

**Figure 3.** The consistency of NEOWISE model fits relative to fits based on the SpeX data as a function of various orbital parameters. Each point is a different NEOWISE data set. "Consistent" means that models based on the given data set are consistent with results based on SpeX data. "Inconsistent" means that a model was fit to the data set, but the model was inconsistent with the results based on the SpeX data. "No Fit" means that no models could be fit to the data set. The top three panels show the fit consistency as a function of the solar phase angle (units of degrees), Sun–object distance (units of au), and theoretical blackbody temperature of the object at the time of observation (units of K), respectively. The bottom panels show how these quantities are related to each other. Here the fit consistency is given by the color. The bottom left panel shows solar phase angle and Sun–object distance; the bottom right panel shows theoretical blackbody temperature and Sun–object distance.

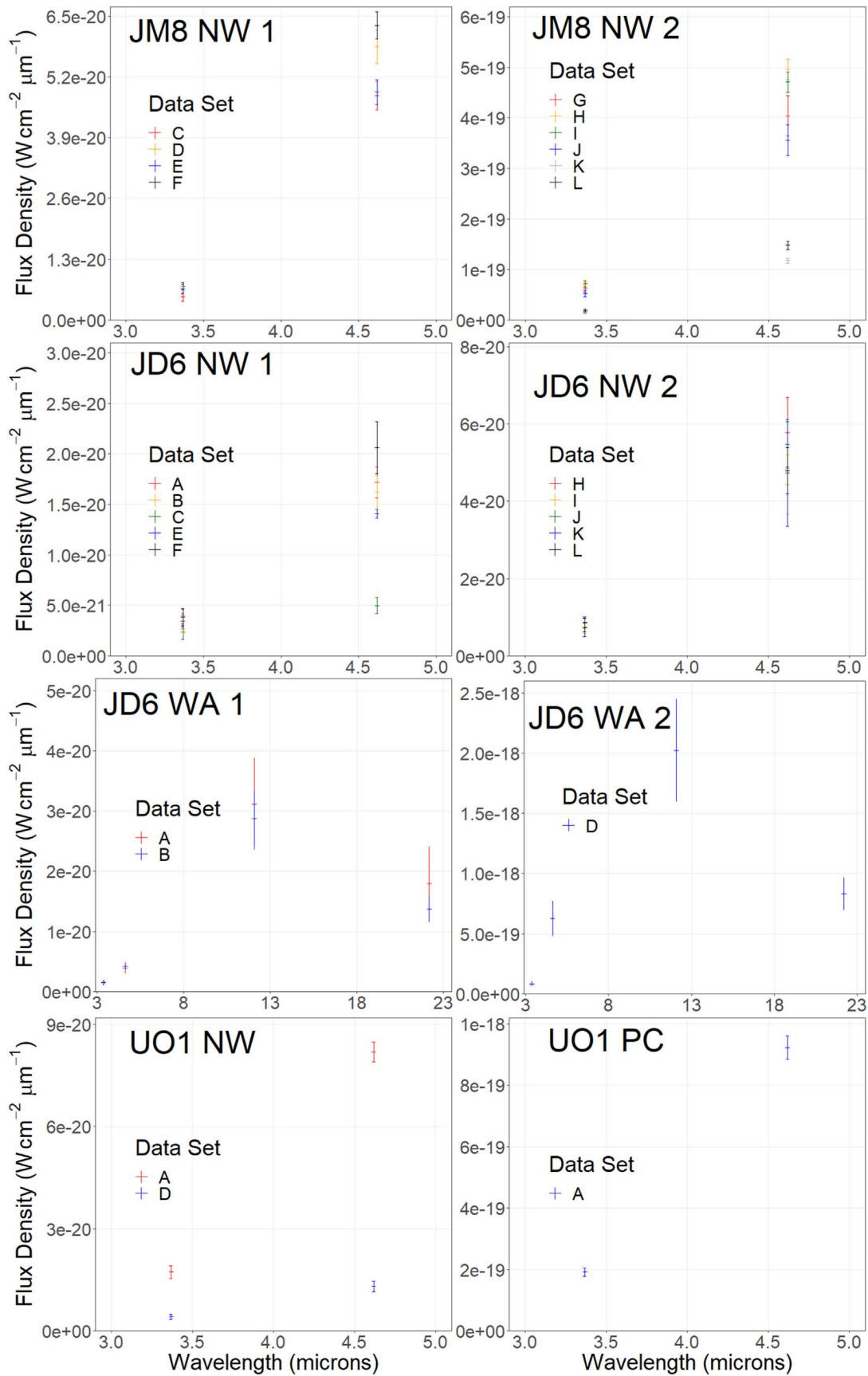

**Figure 4.** Final, averaged NEOWISE data sets. Each panel shows different data sets from different mission phases. WA denotes WISE data, PC denotes post-cryo data, and NW denotes NEOWISE data (Section 2.2). Some panels are further separated for clarity. Letters within each panel denote different data sets. Missing letters indicate sets that were rejected after looking at the raw NEOWISE images (Section 2.2). The y-axis is absolute flux density. The x-axis is wavelength in microns. Note that different panels have different axis scales.

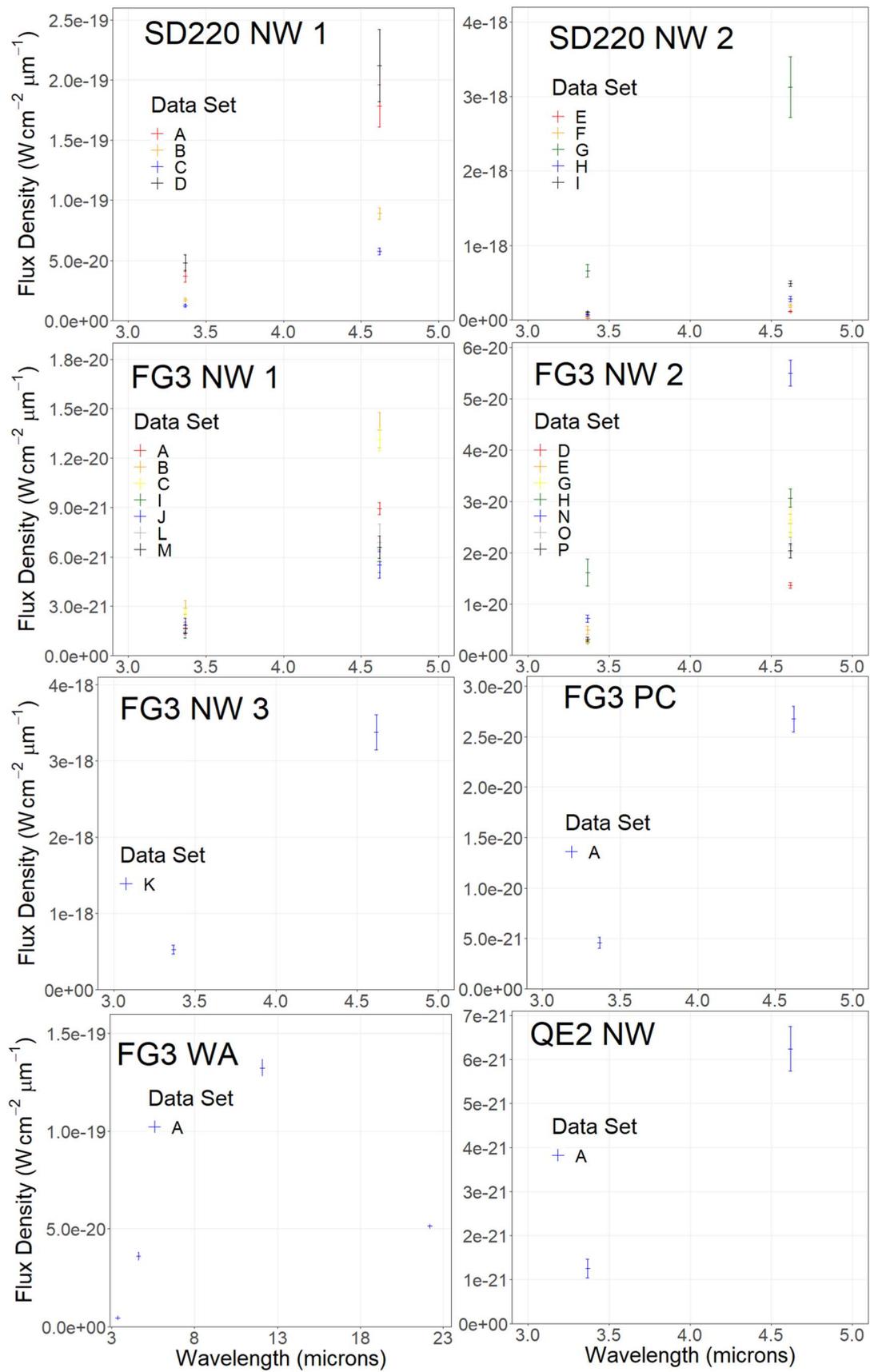

**Figure 4.** (Continued.)

Table 3
Object Background Information and Fixed Model Inputs for Each Object

| Object | Spectral Type | Binary? | Shape | Rotation | Rotation (hr) | H Magnitude | IR Reflectance |
|---|---|---|---|---|---|---|---|
| (53319) 1999 JM8 | X, C | N | Irregular | NPA | 168 | 15.2 | 1.85 |
| (85989) 1999 JD6 | L, K | N | Contact Binary | Retrograde | 7.67 | 17.1 | 1.24 |
| (137032) 1998 UO1 | S | N | Spheroidal | Retrograde | 2.9 | 16.7 | 1.30 |
| (163899) 2003 SD220 | S | N | Elongated | NPA | 285 | 17.2 | 1.60 |
| (175706) 1996 FG3 | C | Y | Spheroidal | Retrograde | 3.594 | 17.8 | 1.47 |
| (285263) 1998 QE2 | Xk | Y | Spheroidal | Prograde | 4.749 | 17.3 | 1.13 |

**Note.** "Binary" indicates whether the object has a separated secondary. "Shape" describes the general shape class of the object, as derived from radar data. "Rotation" describes the object's rotation state, where NPA stands for non–principal axis rotator. The remaining columns are direct inputs to the NEATM-like model for each object. Rotation rates are taken from L. A. Benner et al. (2002; JM8), D. Polishook & N. Brosch (2008; JD6), Pravec (personal communication; UO1), B. D. Warner (2016; SD220), P. Scheirich & P. Pravec (2009; FG3), and A. Springmann et al. (2014; QE2). H magnitudes are taken from L. A. Benner et al. (2002; JM8), V. Reddy et al. (2012; JD6), S. D. Wolters et al. (2008; UO1), Y. S. Bondarenko et al. (2019; SD220), S. D. Wolters et al. (2011; FG3), and N. A. Moskovitz et al. (2017; QE2). IR reflectance values are calculated from our prism spectra in combination with visible data from M. Hicks et al. (2013) when available.

between model results based on the SpeX data and model results based on the NEOWISE data for these objects (Section 4). Thus, issues with phase angle correction at high phase angles are not driving the model inconsistencies. Furthermore, no strong trend with phase angle is identified in our final results (Section 5). Therefore, these differences between our NEATM-like model and NEATM do not affect the results presented here.

Like NEATM, our NEATM-like model outputs predicted thermal and reflected fluxes that are combined into a model spectrum of the total flux from visible to near-IR wavelengths. We run models for a range of input parameters and find which parameters result in the best fit to the observed spectra (or NEOWISE magnitudes), rather than fitting single parameter values to each data set. In this way, we utilize a forward modeling approach. Three free-floating parameters are used in this process: (1) the visible geometric albedo, measured as a dimensionless value from 0.0 to 1.0; (2) the thermal inertia, measured as a positive number in units of $J\,m^{-2}\,s^{-1/2}\,K^{-1}$, hereafter referred to as TIU for thermal inertia unit; and (3) the beaming parameter, a dimensionless scaling factor that generally ranges between ∼0.5 and 2.5. The beaming parameter adjusts the differences between the observed and predicted flux to provide a first-order accounting for the model assumptions. A beaming parameter of 1.0 would therefore mean a model where no additional scaling is done, while nonunitary values would allow scaling to adjust for effects such as surface roughness, nonspherical shapes, local shadowing, or high obliquities. Together, these three parameters are varied for each data set to find the model spectra that best fit the data. For more details on the fitting procedure used here, see S. A. Myers et al. (2023).

The model takes as fixed inputs the rotation period, a visible-to-near-IR reflectance ratio, Earth–object and Sun–object distances, solar phase angle, emissivity, and H magnitude. By taking H magnitude as an input and fitting a visible geometric albedo, we can then calculate a model diameter. This allows us to compare our model diameters with diameters modeled or measured using other methods. The H magnitude values we use are taken from the published literature when available or from JPL Horizons. For elongated objects (JM8, JD6, and SD220) with a range of possible effective diameters, depending on viewing geometries, models are run with multiple input H magnitudes for the NEOWISE data. (This is not required for the SpeX data, which are relative flux and thus have the effects of H magnitude scaled out.) These H magnitude ranges are calculated using the diameter ranges of interest and the best-fitting albedos to the SpeX data. We only compare the diameters we calculate using the H magnitude and model albedo with published diameters when using H magnitudes taken from the literature. Thus, we only compare our calculated diameters with literature diameters when working with the nominal H magnitudes for an object, not the extended ranges. This is done to avoid circularity, as it is the H magnitude that we vary to account for elongated objects.

For all objects the rotation period is taken from available light-curve and radar measurements. The visible-to-near-IR reflectance ratio, measured at 2.5 μm, is estimated from our SpeX prism spectra or measured from visible spectral data when available. This value is a correction factor used to relate the visible albedo to the near-infrared albedo. Orbital ephemerides are calculated for each observation using ESA NEODyS-2.[5] The emissivity is generally set to 0.9, in accordance with typical assumptions for NEA emissivity values (L. A. Lebofsky & J. R. Spencer 1989). However, for the JD6 WISE data, some models were run with an emissivity of 0.7. This was done to investigate whether emissivities of 0.7 might be a better assumption at W3 wavelengths for WISE data (J. Moeyens et al. 2020). JD6 was chosen for this analysis, as it had the most complete and varied WISE data available of all objects investigated in this paper. A list of all model inputs used for each object is shown in Tables 1, 2, and 3.

### 3.2. Model Fitting

For each data set, a range of NEATM-like models are generated across a range of visible albedos, thermal inertias, and beaming parameters. Each model is then compared with the data set using an objective function (see below). The goal of this analysis is to identify the ranges of visible albedos and thermal inertias that can be said to fit each data set independently. We do not attempt to identify the beaming parameter ranges because beaming parameter depends strongly on viewing geometry. Therefore, we expect that the beaming parameters that can fit a given data set will change even across a single night of observations. This is because the beaming parameter is a nonphysical scaling factor that accounts for the assumptions inherent in the model. Thus, we are interested in the aggregate ranges of beaming parameters required to produce well-fitting spectra for each object as a whole, but we do not consider the beaming parameter a fixed property that we are attempting to infer for every data set. For each data set we therefore seek to identify only the ranges of visible albedo and thermal inertia that produce model spectra consistent with the observations.

---

[5] https://newton.spacedys.com/neodys/index.php?pc=3.0

First, for the SpeX data, we calculate a reduced $\chi^2$ between the models and the data. In general, this calculation is only done for the data between 3.0 and 4.05 $\mu$m because this is the part of the spectrum with the strongest thermal emission and little overlap with atmospheric water vapor lines. However, for some post-upgrade data a range of 3.5–5.05 $\mu$m is used. It is important to note that this value is not a formal $\chi^2$ because the uncertainties in the data are dominated by systematic effects and the spectral data points are not independent. Thus, our reduced $\chi^2$ does not reach a minimum value of unity. Furthermore, this means that $\chi^2$ values for different data sets are not directly comparable. For more discussion on the methods used for analyzing the $\chi^2$ values for these data, see S. A. Myers et al. (2023).

Instead, our reduced $\chi^2$ value is used to define a cutoff between models that fit the data and models that do not. This cutoff is reasonable because as the model parameters change monotonically the models move monotonically relative to the data. This is true of all three free-floating parameters: visible albedo, thermal inertia, and beaming parameter (Figure 5). Thus, assuming adequate beaming parameter coverage, all models in the parameter ranges below the cutoff are guaranteed to fit the data. Here we define a model fit as any model that falls within the 1$\sigma$ uncertainties of the data (Figure 6). The same uncertainty range (1$\sigma$) is used for all data sets to ensure similar comparisons between all data sets. Models are also checked to ensure that they are fitting the thermal upturn region ($\sim$2.0–2.5 $\mu$m) and the thermally dominated region ($\gtrsim$3.0 $\mu$m). Inconsistencies are discussed in more detail for individual objects below.

For the NEOWISE data, due to the small number of data points, we can directly check whether or not the models fall within the uncertainties of the particular bands. Thus, the $\chi^2$ calculation is not required. However, the end result is the same: a list of models that fit the data versus a list of models that do not.

After the cutoff has been applied, the model ranges that fit a given data set can be plotted as a $\chi^2$ map (Figure 7, left panel). In this example, every colored square represents a different model that can be said to fit the data, with good fitting to poor fitting corresponding to purple to red. Once a similar map has been generated for every object for a given data set, they can be overlaid to create a final fit map for the object (Figure 7, right panel). In this example, all the $\chi^2$ maps for each SpeX data set for UO1 have been overlaid. The fit map now reports the number of data sets for which a given model fits the data. The more data sets that can be described by a given model, the more confident we can be that the model represents the object's overall properties, not just as it appears at one particular moment. This process is the same for the NEOWISE data, except the individual data setplots indicate whether the models fit or do not fit the data rather than a $\chi^2$ value.

We note that, especially for the NEOWISE data, not all data sets for a given object necessarily have models that overlap, or have models that can describe the data. In Figure 8 we show both the NEOWISE observations for SD220 and the final fit map. The three omitted data sets (B, C, E) have no models that fit the data. Out of the remaining six data sets that do have fitting models, the models that fit do not necessarily overlap. Therefore, the maximum number of overlapping data sets (here four) is far less than the total number of data sets examined (nine). Thus, there are many situations where models based on the NEOWISE data fail to produce consistent results.

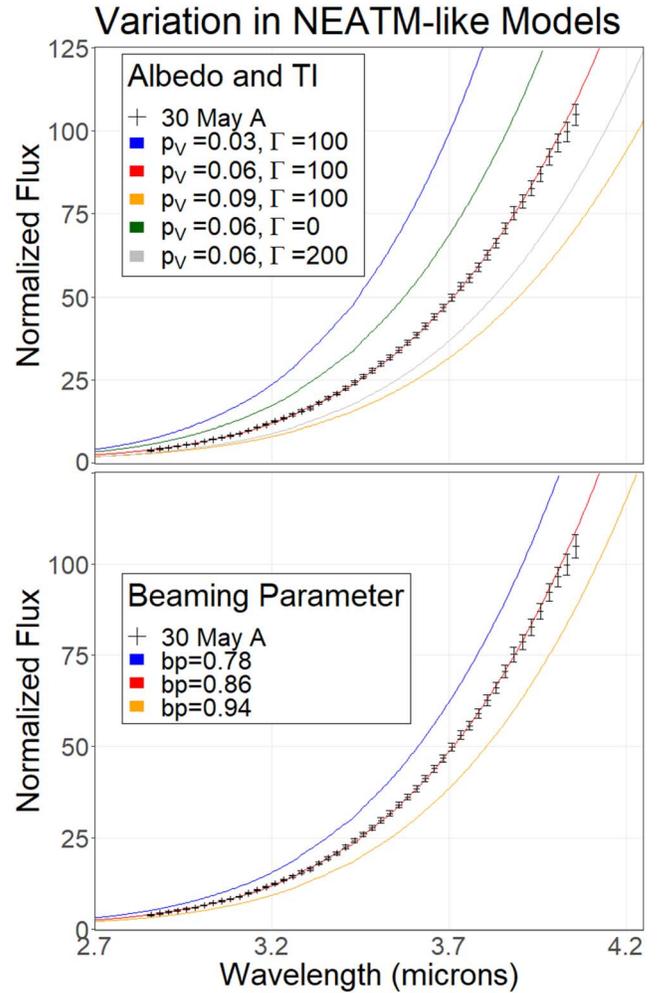

**Figure 5.** Various NEATM-like model fits to a sample SpeX data set for QE2. The different colored lines show different models. Here the y-axis is normalized flux, normalized to 1.0 at $\sim$1.6 $\mu$m. The x-axis is wavelength in microns. As the model parameters change monotonically, the models move monotonically across the data set. This property of the model allows us to easily identify a range of model parameters that can be said to fit the data. Top: the beaming parameter is held at 0.86, and the visible albedo and thermal inertia are allowed to vary. Bottom: the albedo is held at 0.06, thermal inertia is held at 100 TIU, and the beaming parameter is allowed to vary.

Finally, the free-parameter model ranges explored are set to try to fully explore the fit space for a given object. However, values of visible albedo and thermal inertia are limited to exclude physically implausible values. Thus, some fit spaces are technically unconstrained on the upper bound in some instances.

## 4. Model Results

The above procedure was carried out for every data set for each of our six objects. Our six objects were chosen based on numerous criteria, including that they all had readily available data from both SpeX and NEOWISE, preferably over many apparitions. In addition, they have been observed with radar, providing robust size and rotation estimates, and they represent a range of characteristics, including spectral type, shape, and observed solar distances. Furthermore, many of our targets are classified as potentially hazardous asteroids (PHAs) and thus are of interest owing to the potential threats they could pose to Earth. All data used are shown in Figures 1 and 4, and a summary of all observations is given in Tables 1 and 2. Object-specific background information and object-specific model inputs are listed in Table 3. Below we give a brief background of each object and present the modeling results.

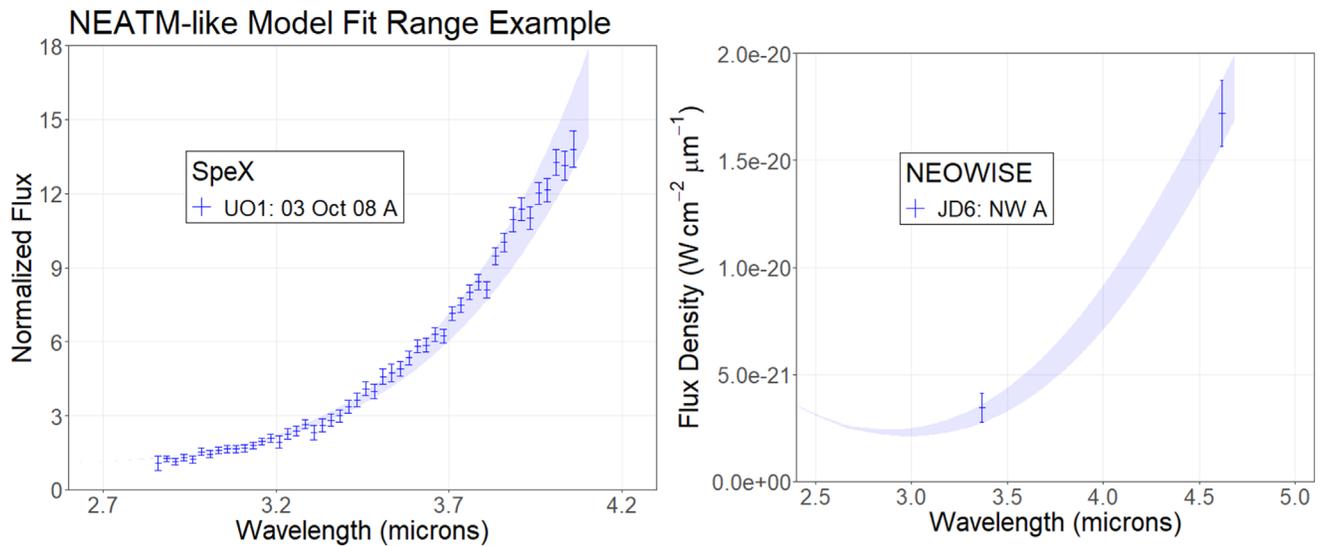

**Figure 6.** Example of the acceptable fit range of our NEATM-like model to the data. Any models that fall within the 1$\sigma$ uncertainties of the data are accepted as fitting the data. Left: a single SpeX data set for UO1. The *y*-axis is normalized flux, normalized to 1.0 at ∼1.6 $\mu$m. The *x*-axis is wavelength in microns. Right: a single NEOWISE data set for JD6. The *y*-axis is absolute flux density. The *x*-axis is wavelength in microns.

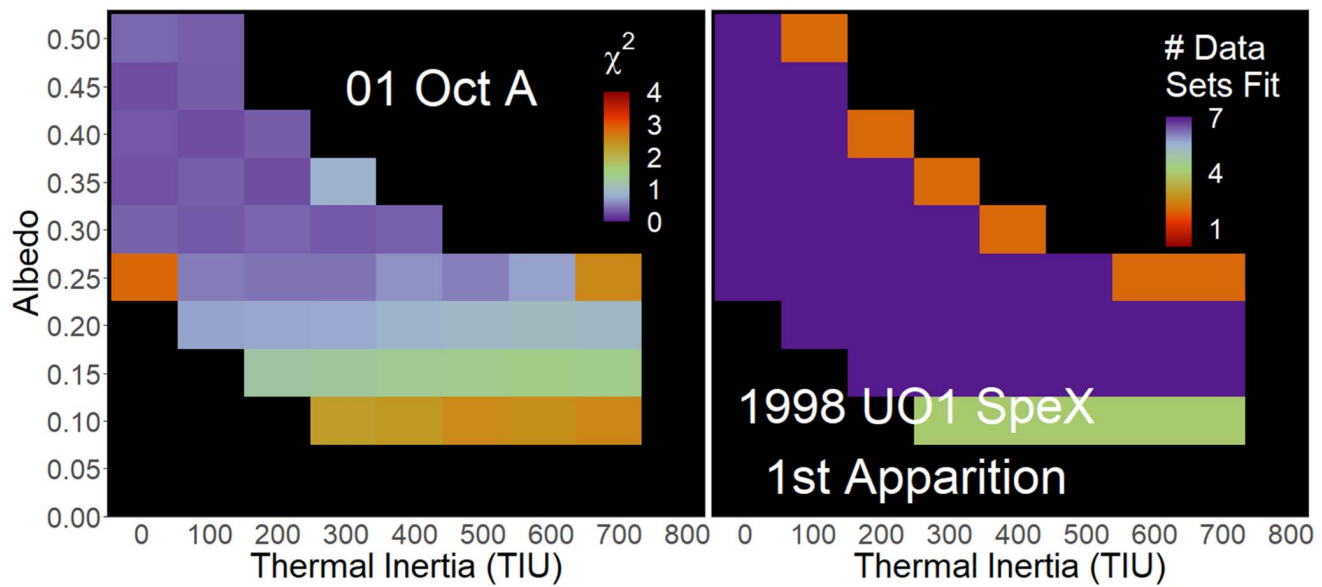

**Figure 7.** Left: example $\chi^2$ fit map for model fits to a single data set (2008 October 1 A) for UO1. Every colored (i.e., nonblack) square represents a different model that can be said to fit the data, with good fitting to poor fitting corresponding to purple to red. Right: example final fit map. Here we show the first SpeX apparition of UO1 with all seven total data sets in 2008 (see Table 1). Every colored square represents a different model, with the color indicating the number of data sets from that apparition that can be fit by the given model. Thus, purple squares represent models that fit more data sets, thus indicating a higher likelihood that the given parameters can describe the object's overall properties, not just properties associated with a particular viewing geometry. In both plots the *x*-axis is the thermal inertia in TIU and the *y*-axis is visible albedo. Here thermal inertia was sampled between 0 and 800 TIU in steps of 100 TIU and the visible albedo was sampled between 0 and 0.5 in steps of 0.05.

When analyzing the model results, we typically follow the same scheme for each object. The SpeX data sets that make up each apparition are examined as a group. For the NEOWISE data, we examine all the NEOWISE and post-cryo data sets together. For these we separately identify models consistent with both the W1 and W2 data points and models consistent with the W2 data point only. A similar scheme is used for objects with WISE data. Models are identified that are consistent with all four bands; with bands W2, W3, and W4; with bands W3 and W4; with bands W2 and W3; with bands W1 and W2; and with band W2 alone. In this way we are able to examine how the different NEOWISE bands affect consistency with the SpeX results. In all cases results are shown for band combinations that had any models that met the criteria. For example, if no plots are shown for models to the W3 and W4 bands, that means that no models were found that could fit both of those bands simultaneously.

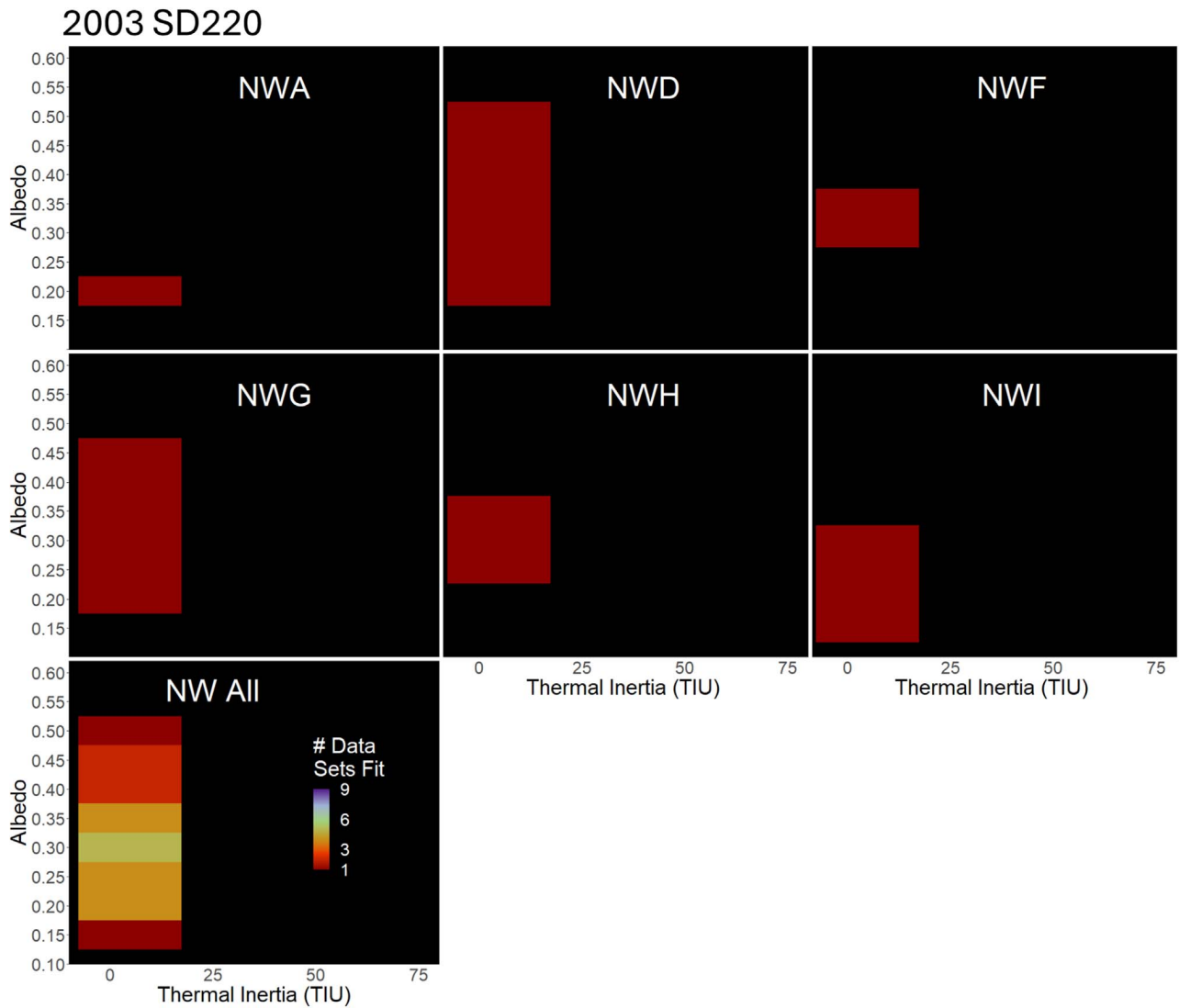

**Figure 8.** Examples of how all data sets for a given object do not necessarily overlap. Here we show NEOWISE data for SD220. The bottom left panel shows the overlap of all NEOWISE data sets for SD220. The other panels show individual data sets, where colored squares indicate models that fit the data. The three omitted data sets (B, C, E) have no models that produce spectra consistent with the data. Furthermore, we note that not all data sets are consistent with each other. Thus, there are only six data sets with fitting models, and the maximum number of overlapping data sets (four) is less than the total number of data sets (nine). This can be seen in the bottom left panel. In other words, out of all nine data sets, the model with the most overlap only fits four data sets. In all plots the *x*-axis is the thermal inertia in TIU and the *y*-axis is visible albedo. We note that for SD220 we only examine a thermal inertia of 0 TIU, due its extremely slow rotation (Section 4.4); however, for all other objects a range of thermal inertias are investigated.

### 4.1. (53319) 1999 JM8

JM8 is a dark, irregularly shaped NEA that is classified as a P-type object in the Tholen taxonomy (D. J. Tholen & M. A. Barucci 1989). It is a long-period, non–principal axis rotator (L. A. Benner et al. 2002). Furthermore, it is one of the largest known NEAs, with a diameter of ∼7 km (L. A. Benner et al. 2002). As a large object and PHA, it has been the subject of numerous previous investigations, including radar observations from both Arecibo and Goldstone (L. A. Benner et al. 2002). Previous authors have reported an albedo range of 0.01–0.05, *H* magnitudes of ∼15–16.5, and *G* values of approximately −0.1 to 0.05 (L. A. Benner et al. 2002; V. Reddy et al. 2012; R. P. Binzel et al. 2019; J. R. Masiero et al. 2020). Based on these results, we use free-parameter model ranges of 0.01–0.10 in increments of 0.01 for the visible albedo and 0–100 TIU in increments of 25 TIU for the thermal inertia. Low thermal inertia values are chosen because, as a long-period, non–principal axis rotator, JM8's surface is unlikely to be strongly affected by thermal inertia, as the surface has ample time to come into thermal equilibrium with the solar flux. Two beaming parameter ranges are investigated, a lower range of 0.70–1.20 in increments of 0.01 and an upper range of 0.70–1.50 in increments of 0.01.

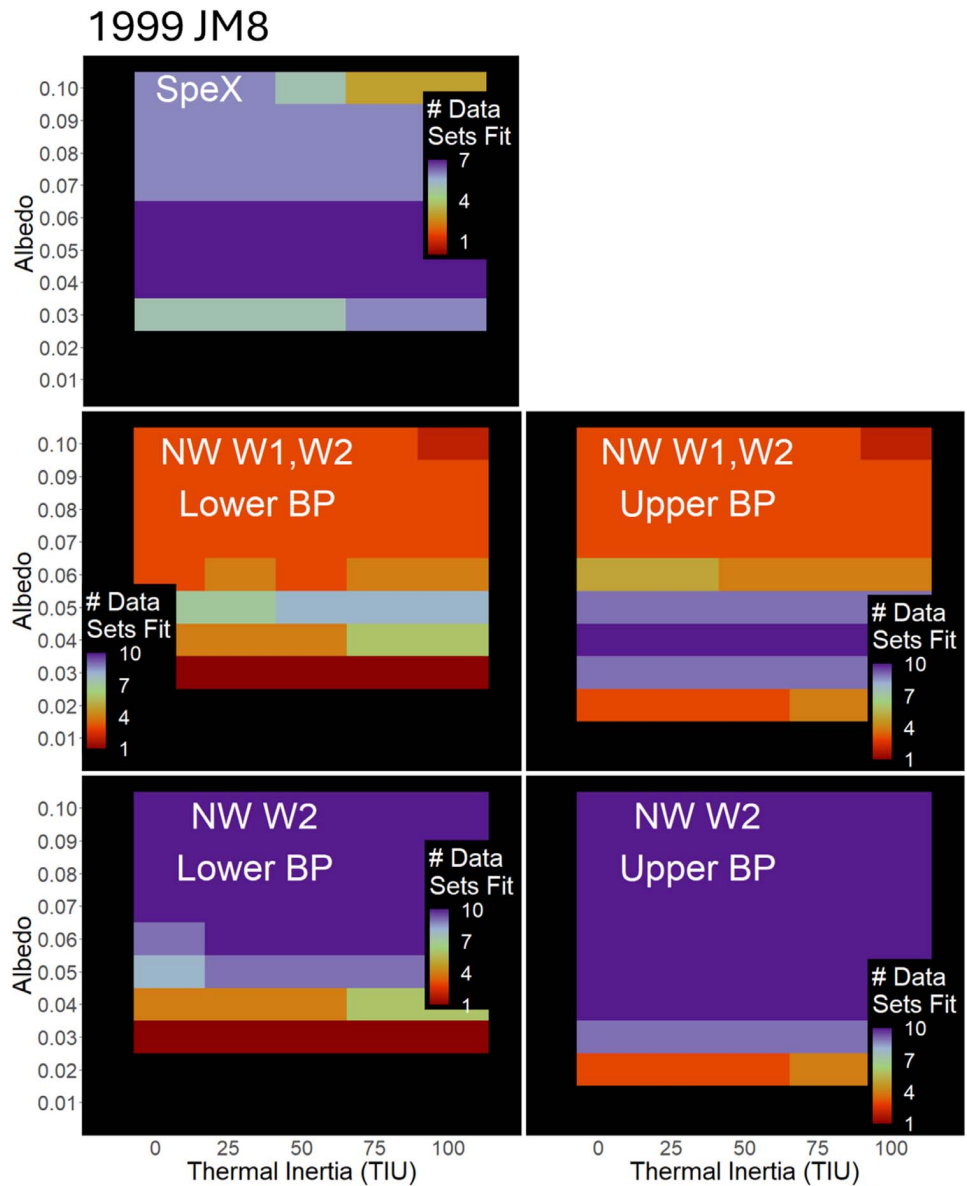

**Figure 9.** Final fit maps for JM8. (For an example of how to read these fit maps, see Figure 7.) The panels show, from left to right and top to bottom, the single SpeX apparition; the NEOWISE data with both bands W1 and W2, first looking at the lower beaming parameter range and then looking at the upper beaming parameter range; and the NEOWISE data with band W2 only, first looking at the lower beaming parameter range and then looking at the upper beaming range. The lower beaming parameter range goes from 0.70 to 1.20 in increments of 0.01. The upper beaming range extends to 1.50. The SpeX beaming parameter range goes from 0.70 to 1.20 in increments of 0.01. There are seven total SpeX data sets and 10 total NEOWISE data sets. In all plots the x-axis is the thermal inertia in TIU and the y-axis is visible albedo.

Only one apparition of SpeX data is available, composed of seven individual data sets. We also fit 10 NEOWISE data sets. Final fit maps are shown in Figure 9. The SpeX data are all consistent with each other, restricting the visible albedo between 0.04 and 0.06 and confirming that there is little dependence on thermal inertia. These visible albedo results are overlapping with, but slightly higher than, previous estimates for JM8 (0.01–0.05). Using these inferred albedo values with previously published $H$ magnitude values gives a diameter of ∼5–6 km, slightly smaller than the radar measurement of 7 km. These slight differences may be due to the increased salience of effects such as self-shadowing at the relatively high phase angles at which we observed JM8, especially for such low albedos.

We also note that there are inconsistencies between the models that fit the thermal upturn region and those that fit the thermally dominated region. For three of our four nights of observations, 2008 May 8, 10, and 15, fits specifically to the thermal upturn region only constrain the visible albedo to be greater than 0.01. The fourth night, 2008 May 13, was unable to produce any models consistent with the thermal upturn region, with the models being cooler than the data. Visible albedos in this lower range are more consistent with previous studies and with the radar observations, implying that the thermally dominated region may be "too cold" relative to what is expected, possibly due to the extremely dark nature of this object. Furthermore, simple models like our NEATM-like model do not account for changes in the slope of the reflected spectra, which may be affecting the model results.

The NEOWISE data show a similar pattern to the SpeX data, and all 10 NEOWISE data sets are consistent with the SpeX data. We do note that there are very few models that fit every NEOWISE set, particularly at the lower beaming parameter range, as can be seen by the banding effect in the NW W1 and W2 panels of Figure 9. Furthermore, the W2-only fits, although consistent with the SpeX results, do push toward higher visible albedos.

Due to JM8's irregular shape, models for the NEOWISE data were also run with an $H$ magnitude of 14.6 to reflect an effective diameter of 5 km. The upper beaming parameter range was used for these fits. Fits with this smaller diameter are largely similar to the fits with the larger diameter. Small increases in the albedo (∼0.02) are observed. However, in general these fits are less consistent with the SpeX data. One data set, NWC, switches from consistent to inconsistent at the smaller diameter. Furthermore, for fits to both the W1 and W2 bands, the maximum number of overlapping data sets is only 7, as opposed to a maximum of 10 for the nominal diameter. The W2-only fits are restricted to albedos of 0.06 or above, which limits them to the upper range of SpeX consistency.

### 4.2. (85989) 1999 JD6

JD6 is a PHA with an elongated, nonspherical shape. Its taxonomy has been described alternatively as an L or K type (R. P. Binzel et al. 2001, 2019), and it is a contact binary with two, roughly equal-sized lobes (S. E. Marshall et al. 2015). As a PHA, JD6 has been the subject of numerous previous studies, including a reported positive YORP detection (J. Tian et al. 2022) and radar observations (S. E. Marshall et al. 2015). Previous authors have found an albedo range of ∼0.01–0.30, diameters of ∼0.6–3 km, and $H$ magnitudes of ∼16.6–17.1 (E. S. Howell et al. 2008; D. Polishook & N. Brosch 2008; H. Campins et al. 2009; C. A. Thomas et al. 2011; V. Reddy et al. 2012; A. Mainzer et al. 2014b; C. R. Nugent et al. 2016; D. Kuroda et al. 2021). Based on these results, we adopt free-parameter model ranges of 0.01–0.31 in increments of 0.02 for the visible albedo and 0–600 TIU in increments of 100 TIU for the thermal inertia. Two beaming parameter ranges are investigated, an upper one from 0.70 to 1.70 in increments of 0.02 and a lower one from 0.50 to 1.70 in increments of 0.02. The WISE data are investigated with a beaming parameter range of 0.50–2.50 in increments of 0.05.

For JD6 we fit two apparitions of SpeX data, one consisting of five data sets and one consisting of three data sets. (We note that one set of SpeX data, 2015 July 21, was provided to us by V. Reddy (personal communication) and has LXD short data only. Data were reprocessed by us similar to other SpeX data sets.) Ten NEOWISE data sets are fit, as well as three WISE data sets. In addition to the standard fitting procedures described above, we also fit the WISE data in bands W3 and W4 with models where the emissivity is set to 0.7. This was done to investigate whether or not lower emissivity values are a better assumption at W3 wavelengths (J. Moeyens et al. 2020). As such, these models were examined for all band combinations except the W1 and W2 band together and the W2 band by itself. (Models were only fit to band combinations that included the longer-wavelength bands where the lower emissivity may be important.) JD6 was chosen for this analysis because it is the only object in our data set that was observed by WISE across more than one apparition and had usable data in all four bands. Final fit maps are shown in Figures 10 and 11.

All SpeX data are generally consistent with each other. Large fit spaces are identified for each apparition separately, and the fit spaces for the two apparitions largely overlap. The models are not tightly constrained, but an albedo range of roughly 0.08–0.31 and a thermal inertia range of roughly 0–600 TIU are found. These ranges are generally consistent with previous work (albedos of 0.01–0.31 and predictions of a low thermal inertia) but do exclude lower-albedo solutions found by some studies (H. Campins et al. 2009; C. A. Thomas et al. 2011; A. Mainzer et al. 2014b; D. Kuroda et al. 2021). Using these inferred visible albedos and previously measured $H$ magnitudes, we get diameters of ∼1.0 to 1.9 km, consistent with radar findings that the maximum extent of the object does not exceed 3 km (S. E. Marshall & E. S. Howell et al. 2017). We also note that observations on 2010 July 22 show a slight discrepancy between the models that fit the thermal upturn region and the thermally dominated region, although these differences are slight enough to not affect the final results. Fits to the upturn region exclusively are not constraining on the albedo.

Although the SpeX data are not very constraining, interesting comparisons can be made with the NEOWISE data. Only 1 of the 10 available NEOWISE data sets did not have any fitting models (C). The remaining NEOWISE data sets were all consistent with the SpeX results when looking at both bands together and the W2 band by itself. However, we do note that (1) the NEOWISE data are not all consistent with each other and (2) the fits to the W2 band by itself do tend to push the visible albedos higher than for the SpeX data. No models are found that fit the W1 and W2 bands together for WAB; however, the fits based on the other two WISE data sets are consistent with the SpeX data. Fits to the W2 band alone produce consistent but almost entirely unconstrained fits.

The WISE data were examined looking at emissivities of both 0.90 and 0.70. These results are shown in Figure 11. Here we can see that well-constrained fits are possible for both emissivities in all band combinations examined. All fits, except for the fits to all four bands at emissivity of 0.70, are consistent with the SpeX data. Thus, in general, we find that using a different emissivity value for the longer-wavelength bands does not change the fit consistency in most cases. However, it does in all cases lower the visible albedo estimate. Therefore, we can make no solid conclusion about whether or not a lower emissivity value is a better assumption for longer wavelengths.

The NEOWISE data for JD6 are also fit with different $H$ magnitudes to account for different effective diameters, due to JD6's elongated shape. JD6 has a minimum effective diameter of ∼1.3 km and a maximum extent of ∼3.0 km, based on radar data, corresponding to an $H$ magnitude of 15.3. The above fits were all done with an $H$ magnitude equivalent to a diameter of 1.3 km. Thus, an additional fit was done with an $H$ magnitude of 15.3. For these fits, models could only fit the NEOWISE data when fitting the W2 band by itself. Fits to the WISE data could not be found for all four bands together, or for bands W1 and W2 by themselves. In all other cases where models could be fit the albedos increased to the upper edge of the fit range ($\gtrsim 0.29$). Thus, these fits were still consistent with the SpeX data, but at the opposite end of the fit range. Thus, the expanded diameter ranges have the effect of broadening the fit space of the WISE data to be largely unconstrained, while not affecting the overall consistency of the NEOWISE modeling results with the SpeX data.

### 4.3. (137032) 1998 UO1

UO1 is a rapidly rotating, spherical, S-type NEA (C. A. Thomas et al. 2011). Very little work on UO1 has been published to date. Radar observations from both Goldstone and Arecibo are available for UO1 (S. Marshall et al. 2014). Previous authors have found (largely unconstrained) albedos of around 0.30, diameters of ~1.1 km, and an *H* magnitude of ~16.5 (S. D. Wolters et al. 2008; C. A. Thomas et al. 2011). We ran models of UO1 with free-parameter ranges of 0.05–0.50 in increments of 0.05 for the visible albedo and 0–700 TIU in increments of 100 TIU for the thermal inertia. A beaming parameter range of 0.70–1.20 in increments of 0.01 was used as well.

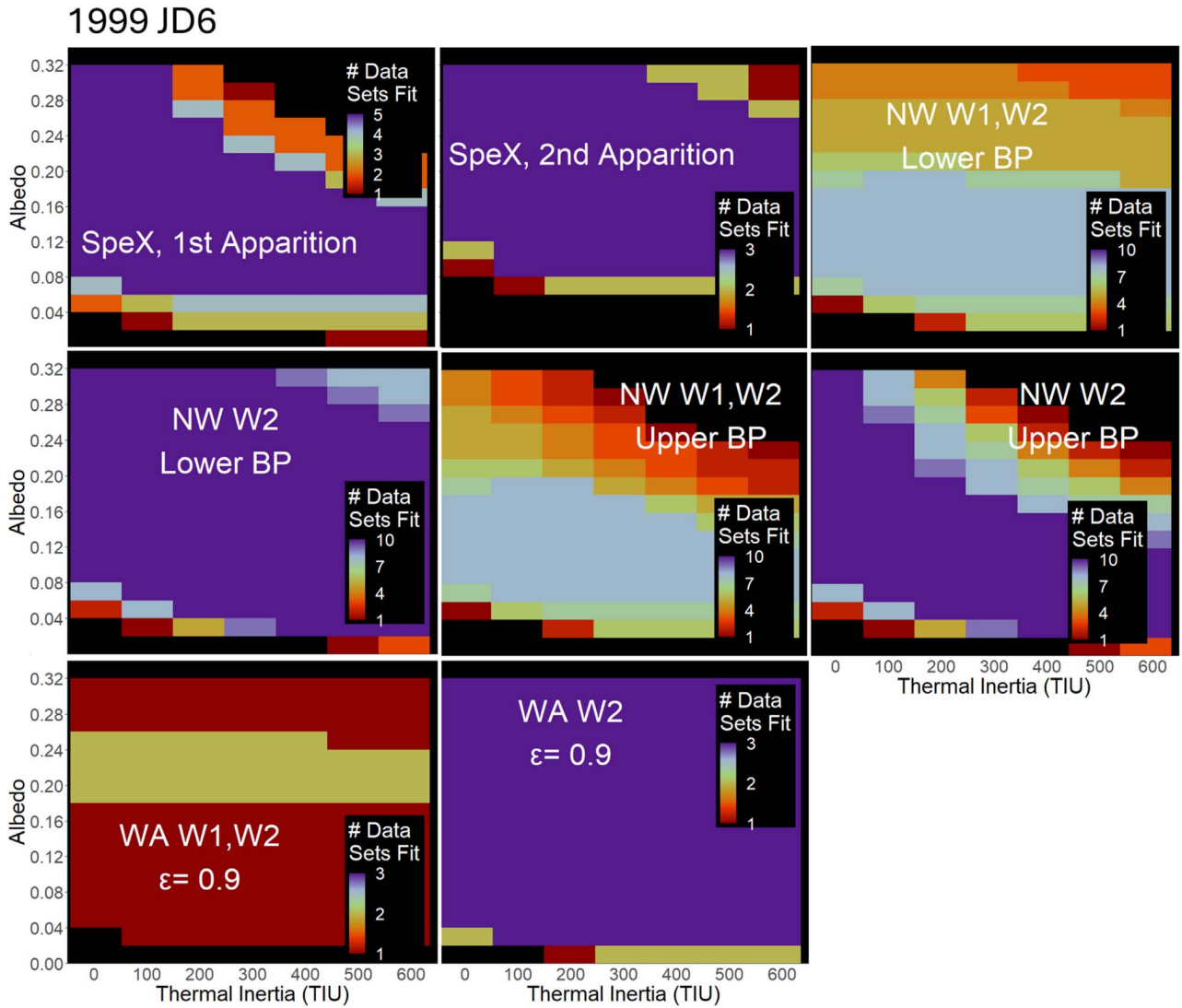

**Figure 10.** Final fit maps for JD6. (For an example of how to read these fit maps, see Figure 7.) The panels show, from left to right and top to bottom, the two SpeX apparitions; the NEOWISE data with the lower beaming parameter range, first looking at both bands W1 and W2 and then looking at band W2 only; the NEOWISE data with the upper beaming parameter range, first looking at both bands W1 and W2 and then looking at band W2 only; and the WISE data with emissivity of 0.9, first looking at both bands W1 and W2 and then looking at band W2 only. The NEOWISE lower beaming range goes from 0.50 to 1.70 in increments of 0.02. The NEOWISE upper beaming parameter range goes from 0.70 to 1.70 in increments of 0.02. The WISE beaming parameter range is 0.50–2.50 in increments of 0.05. The SpeX beaming parameter range goes from 0.70 to 1.70 in increments of 0.02. There are five total SpeX data sets in the first apparition and three total in the second apparition, 10 total NEOWISE data sets, and three total WISE data sets. In all plots the *x*-axis is the thermal inertia in TIU and the *y*-axis is visible albedo.

Two apparitions of SpeX data were used. The first apparition consisted of seven data sets, while the second apparition consisted of eight. Two NEOWISE data sets and one post-cryo data set were also analyzed. Final fit maps are shown in Figure 12. The two SpeX apparitions are largely consistent with each other, with slight differences between them. We note that one night of observations, 2010 October 8, displayed a mismatch between the models that fit the thermally dominated region and the models that fit the thermal upturn region, with the thermal upturn region generally requiring visible albedos lower by 0.10–0.15. However, regardless of which values are used, the results are still consistent with other data sets from that apparition. Overall, we find a visible albedo range of roughly 0.25–0.50 and a thermal inertia range of roughly 0–400 TIU to be consistent with all the SpeX data. This is a slightly higher albedo than one of the previously published estimates of $0.13^{+0.11}_{-0.06}$ (C. A. Thomas et al. 2011). Using these inferred visible albedo values and the published *H* magnitude value gives a diameter of ~0.75–1.7 km, consistent with existing radar observations that find a maximum diameter of 1.2 km (S. Marshall et al. 2014).

For UO1 we see that the NEOWISE data are largely consistent with the SpeX data, although one NEOWISE data set was consistent with only the first SpeX apparition. The W2-only models are consistent with both apparitions in all cases.

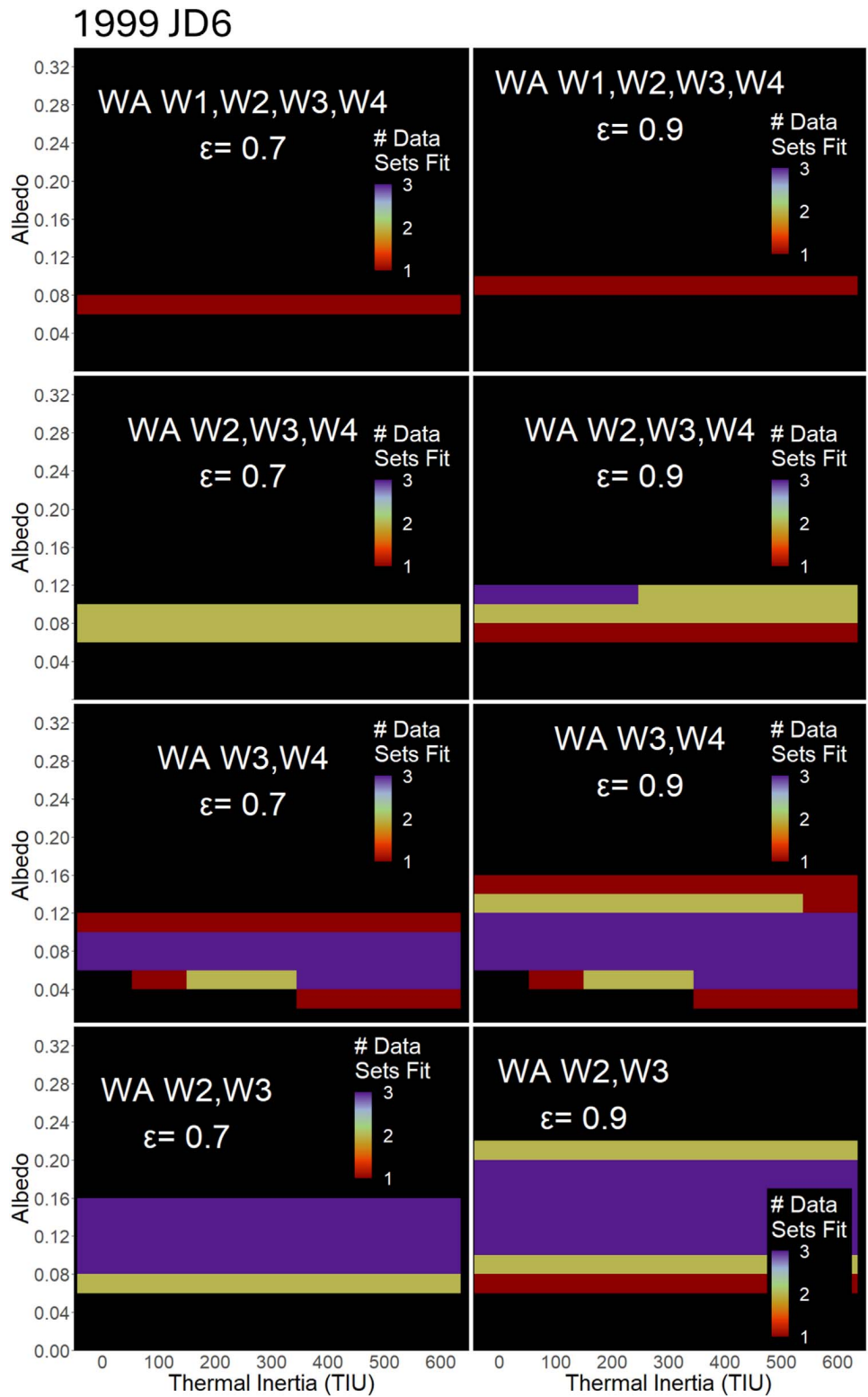

**Figure 11.** Final fit maps for JD6. (For an example of how to read these fit maps, see Figure 7.) Here we show all explored band combinations for the WISE data. The panels show, from left to right and top to bottom, all four WISE bands, first looking at emissivity of 0.7 and then looking at emissivity of 0.9; WISE bands W2, W3, and W4, first looking at emissivity of 0.7 and then looking at emissivity of 0.9; WISE bands W3 and W4, first looking at emissivity of 0.7 and then looking at emissivity of 0.9; and WISE bands W2 and W3, first looking at emissivity of 0.7 and then looking at emissivity of 0.9. (For bands W1 and W2 and band W2 alone, see Figure 10.) In all cases the beaming parameter range is 0.50–2.50 in increments of 0.05. There are three total WISE data sets. In all plots the *x*-axis is the thermal inertia in TIU and the *y*-axis is visible albedo.

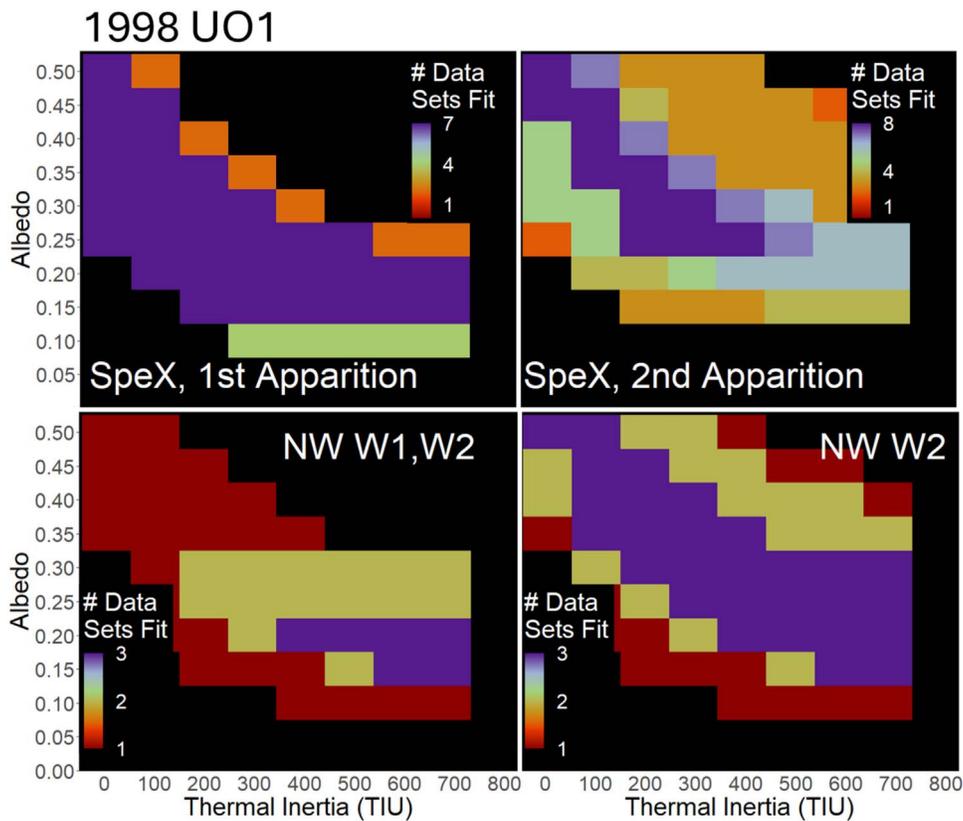

**Figure 12.** Final fit maps for UO1. (For an example of how to read these fit maps, see Figure 7.) The panels show, from left to right and top to bottom, the two SpeX apparitions and the NEOWISE data, first looking at both bands W1 and W2 and then looking at band W2 only. The beaming parameter range goes from 0.70 to 1.20 in increments of 0.01. There are seven total SpeX data sets in the first apparition and eight total in the second apparition, and there are three total NEOWISE data sets. In all plots the x-axis is the thermal inertia in TIU and the y-axis is visible albedo.

### 4.4. (163899) 2003 SD220

SD220 is an S-type, elongated NEA. It is a long-period, non–principal axis rotator and is classified as a PHA and thus has been the subject of a handful of previous studies (Y. S. Bondarenko et al. 2019). Radar observations exist for SD220 from Arecibo, Goldstone, and Canberra (E. Rivera-Valentín et al. 2019; S. Horiuchi et al. 2021). Previous studies have found an albedo of $0.31 \pm 0.041$ (C. R. Nugent et al. 2016); diameters of ∼0.5–3 km (C. R. Nugent et al. 2016; Y. S. Bondarenko et al. 2019; E. Rivera-Valentín et al. 2019), with the long axis being >2.5 km and the intermediate axis being ∼1 km (S. Horiuchi et al. 2021); and an $H$ magnitude of ∼17.2 (C. R. Nugent et al. 2016; S. Horiuchi et al. 2021). Based on these results, we adopt a visible albedo range of 0.05–0.50 in increments of 0.05. For the SpeX data we use a thermal inertia range of 0–75 TIU in increments of 25 TIU. For the NEOWISE data we only investigate thermal inertias of 0 TIU because, as a long-period, non–principal axis rotator, SD220's surface is unlikely to be strongly affected by thermal inertia, as the surface has ample time to come into thermal equilibrium with the solar flux. For all data we use a beaming parameter range of 0.5–2.0 in increments of 0.05.

We have two available apparitions of SpeX data, both consisting of four data sets each. There are also nine available NEOWISE data sets. Final fit maps are shown in Figure 13. The SpeX data are all consistent with each other, both within each apparition and between the two apparitions. These results constrain the visible albedo to be between 0.25 and 0.35 for the first apparition and between 0.25 and 0.45 for the second apparition. Visible albedos in this range generally agree with previous results (albedos of 0.27–0.35). When combined with a previously measured $H$ magnitude, these inferred visible albedos give a diameter of ∼0.7–1.0 km. This value is within the range of ∼0.5–2.5 km for spherical equivalent diameters implied by radar observations (E. Rivera-Valentín et al. 2019; S. Horiuchi et al. 2021). Our modeling results also confirm that thermal inertia plays little to no role in determining the surface properties of SD220.

The NEOWISE results are generally consistent with the SpeX results; however, there is limited self-consistency among the NEOWISE data sets themselves (Figure 8). Furthermore, of the nine available data sets, there are three for which no models could be fit. This may be due to the high phase angles at which all the NEOWISE data were taken. Fits to both the W1 and W2 bands and the W2 band alone were largely consistent with the SpeX data, although the W2-only fits showed a preference for higher visible albedos.

We fit the NEOWISE data with models of different $H$ magnitudes corresponding to different diameters as well, due to SD220's elongated shape. The above fits are done with an $H$ magnitude of 17.2, corresponding to a diameter of ∼1 km. Models are also fit with $H$ magnitudes 18.4 and 14.6, corresponding to diameters of 0.5 and 3.0 km, respectively. These are the minimum apparent diameter and maximum apparent diameter. No models can be fit with a diameter of 3.0 km. However, models can be fit with a diameter of 0.5 km. These results generally show individual data sets to be more consistent with the SpeX data. NWA changes from being inconsistent to being consistent, NWB changes from no fit to inconsistent, and NWE changes from no fit to consistent. Thus, with the smaller diameter, eight of the nine NEOWISE sets are consistent with SpeX. However, these eight data sets only overlap with each other at albedos of 0.15 (Figure 13). Furthermore, in the albedo range consistent with the SpeX data, a maximum of six data sets overlap, compared to four at the nominal diameter. Thus, overall the new fit space is slightly more consistent with SpeX but underestimates the albedo. The W2-only fits overall move the albedos down to 0.25 and above, making them more consistent with the SpeX data.

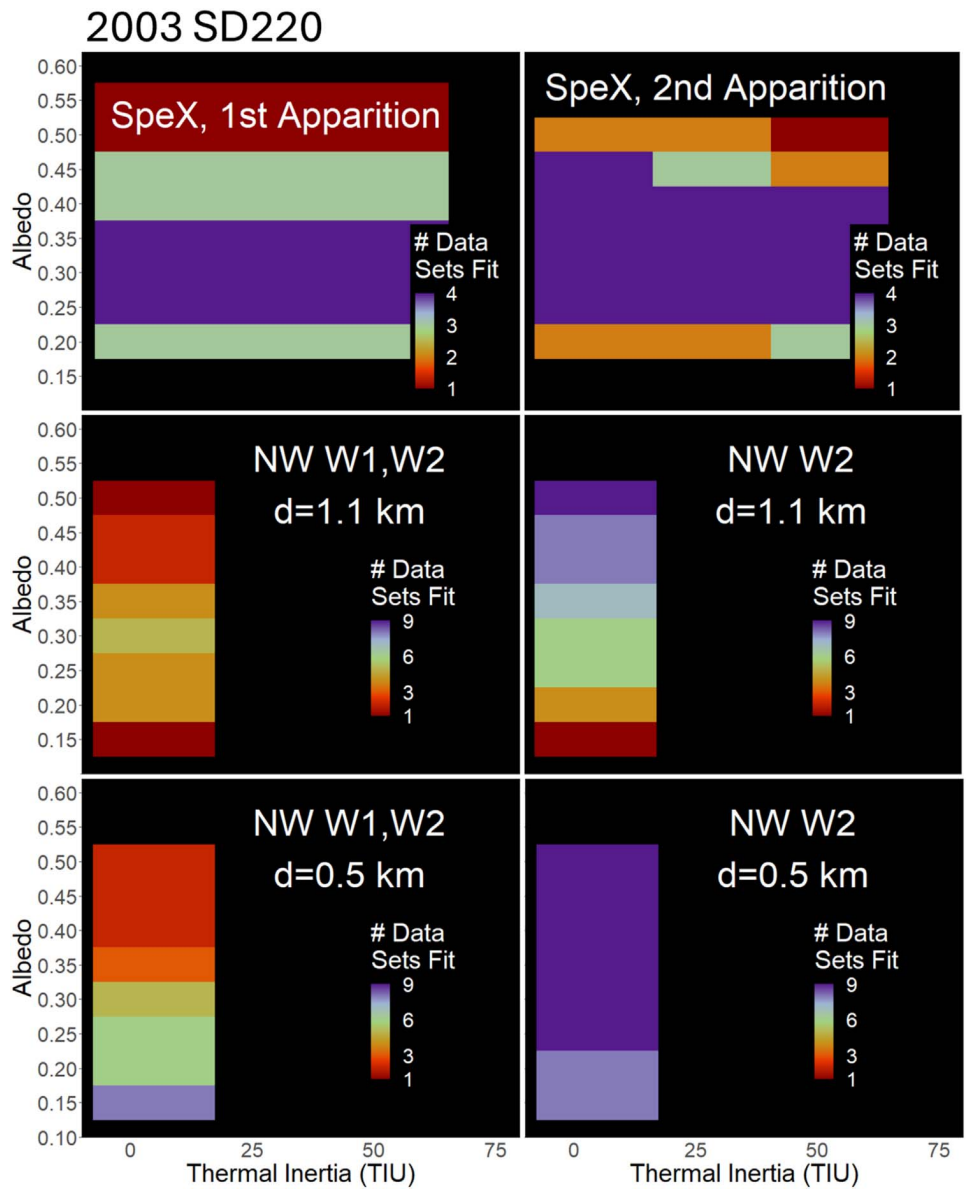

**Figure 13.** Final fit maps for SD220. (For an example of how to read these fit maps, see Figure 7.) The panels show, from left to right and top to bottom, the two SpeX apparitions; the NEOWISE data with the nominal diameter, first looking at both bands W1 and W2 and then looking at band W2 only; and the NEOWISE data with the smaller diameter, first looking at both bands W1 and W2 and then looking at band W2 only. We note that it is the *H* magnitude input to the model that actually changes between these two cases (Section 3). The beaming parameter range goes from 0.50 to 2.00 in increments of 0.05. There are four total SpeX data sets in the first apparition and four total in the second apparition, and there are nine total NEOWISE data sets. In all plots the *x*-axis is the thermal inertia in TIU and the *y*-axis is visible albedo.

### 4.5. (175706) 1996 FG3

FG3 is a dark, binary, Ch-type NEA with a roughly spherical shape that has been the subject of numerous previous studies owing to its status as a PHA and potential mission target. In addition to having been modeled with a variety of thermal models based on ground-based data, FG3 has also been observed with both Arecibo and Goldstone (L. A. Benner et al. 2012). Previous authors have found an albedo range of $\sim$0.01–0.06 (M. Mueller et al. 2011; S. D. Wolters et al. 2011; K. J. Walsh et al. 2012), thermal inertias of $\sim$40–170 TIU (S. D. Wolters et al. 2011; L. Yu et al. 2014), diameters of $\sim$1.5–2.4 km (P. Pravec et al. 2000; M. Mueller et al. 2011; S. D. Wolters et al. 2011; L. A. Benner et al. 2012; K. J. Walsh et al. 2012; L. Yu et al. 2014), *H* magnitudes of $\sim$17.8–18.2, and *G* values of approximately $-0.04$ (M. Mueller et al. 2011; S. D. Wolters et al. 2011). Based on these results, we adopt model free-parameter ranges of 0.01–0.15 in increments of 0.01 for the visible albedo and 0–700 TIU in increments of 100

TIU for the thermal inertia. We explore two different beaming parameter ranges. We explore a lower range of 0.70–1.20 in increments of 0.01 and an upper range of 1.21–1.70 in increments of 0.01.

Overall, for FG3 we fit two apparitions of SpeX data, one made up of 11 data sets and another made up of seven data sets. We also fit 15 NEOWISE data sets, one post-cryo data set, and one WISE data set. Final fit maps are shown in Figure 14. We see that all data sets within each SpeX apparition are consistent with each other, and the two apparitions as a whole are largely consistent with each other. For the first apparition a model with a visible albedo of 0.04 and thermal inertia of 200 TIU is consistent with every data set. For the second apparition a model with a visible albedo of 0.03 and thermal inertia of 100 or 200 TIU is consistent with every data set. Together, these results are largely consistent with previous work on FG3 (albedos of 0.01–0.06 and thermal inertias of 40–170 TIU). Using the $H$ magnitude from S. D. Wolters et al. (2011) and our inferred visible albedos, we calculate a diameter of $\sim$1.5–2.0 km, also consistent with previous work (diameters of 1.5–2.4 km).

It is known that the secondary was eclipsing the primary during the SpeX observations on 2012 January 4. The observations on 2011 November 28 may have also occurred during an eclipse. We find no notable difference in results for these two data sets relative to the other SpeX data. Furthermore, the flux contribution from the secondary did not exceed $\sim$10%. Thus, these eclipses did not affect our results.

Compared to the SpeX data, the models based on the NEOWISE data sets show less consistency. Models could not be found that fit both bands for 4 of our 16 NEOWISE and post-cryo data sets. Of the remaining 12, only 3 had models that were consistent with the SpeX data, 2 of which (K, O) were only consistent at the higher beaming parameter range. In all cases where the NEOWISE models were inconsistent with the SpeX models, the NEOWISE models found either higher visible albedos or higher thermal inertias. The W2-only models at the higher beaming parameter range are more consistent with the SpeX data. The W2-only models at the lower beaming parameter range are consistent in some instances but still generally require higher albedos. A similar trend is found with the WISE data. No models can fit all four bands simultaneously, and those bands that can be fit generally require either higher albedos or thermal inertias. However, we note that many models were consistent with either the first or second SpeX apparition separately, but not both together. The exception is the higher beaming parameter models consistent with the W2 band alone, which were consistent with all the SpeX data.

### 4.6. (285263) 1998 QE2

QE2 is a dark, binary, Xk-type (Bus–DeMeo taxonomy) NEA with a roughly spherical shape. It is a PHA and thus has been the subject of numerous previous studies, including radar observations from Arecibo (A. Springmann et al. 2014). Previous authors have found an albedo range of $\sim$0.01–0.10 (N. A. Moskovitz et al. 2017; S. K. Fieber-Beyer et al. 2020; S. A. Myers et al. 2023), diameters of $\sim$3 km (A. Springmann et al. 2014), and $H$ magnitudes of $\sim$16.4–17.3 (D. E. Trilling et al. 2010; N. A. Moskovitz et al. 2017). These modeling results of QE2 were previously reported by S. A. Myers et al. (2023); however, we give a brief overview here.

One apparition of SpeX data made up of 20 individual data sets was modeled with a visible albedo range of 0.01–0.17 in increments of 0.01 and a thermal inertia range of 0–550 TIU in increments of 25 TIU. A beaming parameter range of 0.55–0.80 in increments of 0.01 was examined. A single NEOWISE data set was also modeled with the same albedo and thermal inertia ranges and with a beaming parameter range of 0.40–0.95 in increments of 0.01. Final fit maps are shown in Figure 15.

All SpeX data sets are consistent with each other, constraining the visible albedo between 0.04 and 0.10 and the thermal inertia between 0 and 425 TIU. The results found for the SpeX data are consistent with previous work; however, the inferred visible albedo is slightly higher than that in previous studies (albedos of 0.01–0.10). A mismatch is also identified between the inferred visible albedo, published $H$ magnitude, and radar measured diameter. Taking the radar diameter and the inferred visible albedo gives an $H$ magnitude of around 16, lower (brighter) than the published value of 17.3. Further discussion of this discrepancy can be found in S. A. Myers et al. (2023).

Furthermore, the NEOWISE data are not consistent with the SpeX data when considering both bands, with the models that fit the NEOWISE data requiring higher albedos relative to the SpeX data. The W2 data are generally consistent, but across a much wider range of parameter space.

## 5. SpeX and NEOWISE Model Comparisons

Our primary interest is in understanding how the results of our simple, NEATM-like model compare when using SpeX data versus NEOWISE data. Thus, for each object we categorize every NEOWISE data set as either no fit, inconsistent, or consistent. No fit indicates a data set for which no fitting models could be found in the investigated parameter ranges. Inconsistent indicates a data set for which fits were found but those fits did not overlap with the fits based on SpeX observations for that object. Consistent indicates a data set for which fits were found and those fits overlapped with the fits based on SpeX observations for that object. These categorizations for every NEOWISE data set are summarized in Table 2.

In summary, we find that many sets are consistent, but a significant number are not. From a total of 53 NEOWISE sets, 9 ($\sim$17%) can be classified as no fit, 12 ($\sim$23%) can be classified as inconsistent, and 32 ($\sim$60%) can be classified as consistent. We note that all SpeX results, with the possible exception of JM8, are in general agreement with previous studies, including radar-derived sizes, which are highly reliable as long as the range of viewing geometries is broad enough (S. J. Ostro 1985; M. C. Nolan et al. 2013; L. A. Benner et al. 2015). Furthermore, in the case of JM8, the NEOWISE-inferred parameters are more dissimilar from previous results than those inferred from the SpeX data. Therefore, we have high confidence that our SpeX results are closer to the true physical values for these objects. Based on our limited sample size, this suggests that inconsistencies between the SpeX and NEOWISE results may indicate potential inaccuracies in the parameters inferred from NEOWISE data when looking at individual objects. We are therefore interested in investigating how these inconsistencies vary as a function of various object characteristics, with the understanding that the statistical power of these correlations is limited owing to our small sample size.

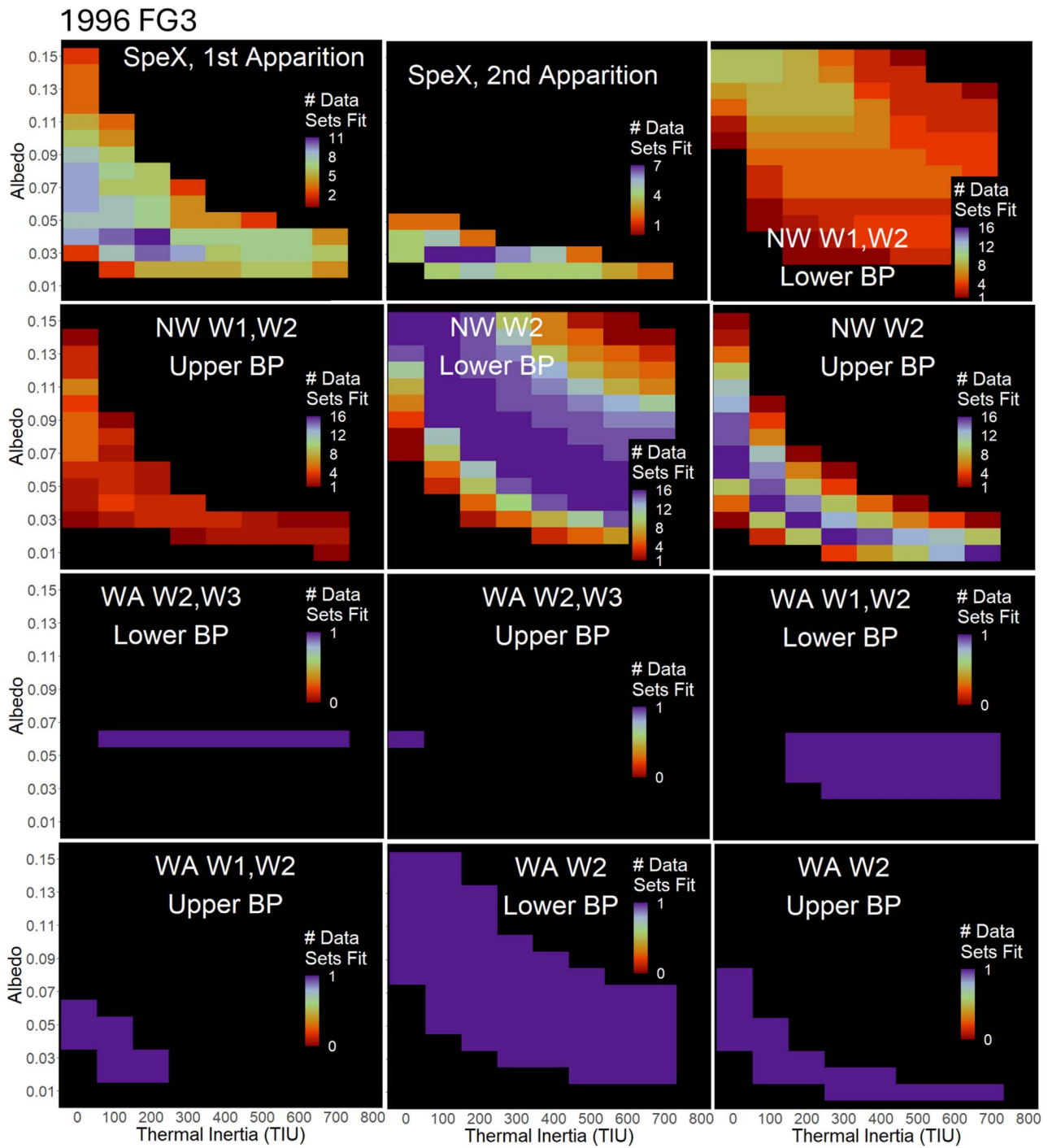

**Figure 14.** Final fit maps for FG3. (For an example of how to read these fit maps, see Figure 7.) The panels show, from left to right and top to bottom, the two SpeX apparitions; the NEOWISE data with both bands W1 and W2, first looking at the lower beaming parameter range and then looking at the upper beaming parameter range; the NEOWISE data with band W2 only, first looking at the lower beaming parameter range and then looking at the upper beaming range; the WISE data with bands W2 and W3, first looking at the lower beaming parameter range and then looking at the upper beaming parameter range; the WISE data with bands W1 and W2, first looking at the lower beaming parameter range and then looking at the upper beaming parameter range; and the WISE data with band W2 only, first looking at the lower beaming parameter range and then looking at the upper beaming parameter range. Note that other combinations of WISE bands did not produce any models that were consistent with the data (Section 4). The lower beaming parameter range goes from 0.70 to 1.20 in increments of 0.01. The upper beaming parameter range goes from 1.21 to 1.70 in increments of 0.01. The SpeX beaming parameter range goes from 0.70 to 1.20 in increments of 0.01. There are 11 total SpeX data sets in the first apparition and seven total in the second apparition, 16 total NEOWISE data sets, and one total WISE data set. In all plots the *x*-axis is the thermal inertia in TIU and the *y*-axis is visible albedo.

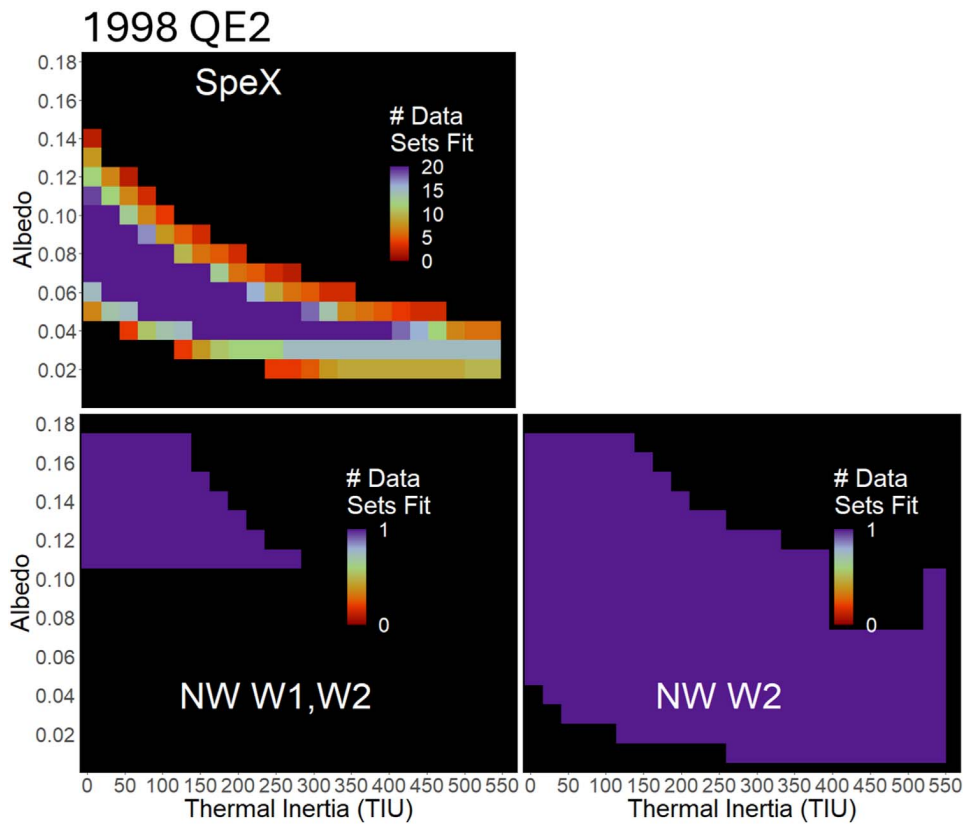

**Figure 15.** Final fit maps for QE2. (For an example of how to read these fit maps, see Figure 7.) The panels show, from left to right and top to bottom, the single SpeX apparition and the NEOWISE data, first looking at both bands W1 and W2 and then looking at band W2 only. The NEOWISE beaming parameter range goes from 0.40 to 0.95 in increments of 0.01. The SpeX beaming parameter range goes from 0.55 to 0.80 in increments of 0.01. There are 20 total SpeX data sets and one total NEOWISE data set. In all plots the *x*-axis is the thermal inertia in TIU and the *y*-axis is visible albedo. (Note that these results were previously presented in S. A. Myers et al. 2023.)

### 5.1. Fit Consistency by Objects and Ephemerides

We first examine how the consistency of the NEOWISE data relative to the SpeX data varies as a function of various viewing geometries and related values. Three key parameters are examined: phase angle, Sun–object distance, and object isothermal blackbody temperature. Figure 3 shows plots of our NEOWISE consistency categories as a function of these parameters. These plots show that there is no apparent correlation between object ephemerides and NEOWISE fit consistency. We also note that all three parameters are highly correlated with each other, as shown in the bottom panels.

In addition to orbital ephemerides, we also examine the consistency of the NEOWISE data with the SpeX data across the different objects looked at here. Two objects, FG3 and QE2, have the majority of NEOWISE data sets characterized as either no fit or inconsistent. For QE2, the one available data set is inconsistent. For FG3, only 3 of the 15 sets are consistent. For the remaining objects, the majority of their NEOWISE data sets ($\gtrsim$70%) are consistent with the SpeX data. We note that for SD220 two sets (A and E) are only consistent using a different $H$ magnitude corresponding to a different effective diameter. However, this is not surprising, due to its highly elongated nature.

The fact that FG3 and QE2 are majority inconsistent or no fit is important because they represent distinct types of objects. FG3 and QE2 are spheroidal and dark objects. They therefore represent objects that have a primitive composition. These objects are also highly spherical and relatively smooth and thus are expected to be reasonably described by the assumptions inherent in our NEATM-like model. It is therefore notable that they are the least consistent. JM8 is also a very dark object in that it is likely composed of primitive, carbonaceous material; however, it does not exhibit the same behavior. Unlike FG3 and QE2, though, JM8 is quite large and therefore has a much brighter absolute magnitude (∼15 compared to ∼17). (JM8 is also a non–principal axis rotator, which may explain some of the difference.)

The other three objects, JD6, UO1, and SD220, are brighter objects in comparison. JD6 is a moderately bright object, belonging to the medium albedo K class, and UO1 and SD220 are S types. UO1 and SD220 in particular have stark differences in their shapes and rotation states, with UO1 being spherical and rapidly rotating and SD220 being an elongated, non–principal axis rotator. The fact that these two objects therefore display similar properties in terms of the consistency of the NEOWISE data with the SpeX data indicates that their relative brightness is likely their relevant common factor. Thus, overall we see that brightness, as a function of both albedo and absolute magnitude, is likely a key factor affecting fit consistency.

We also examine these relationships by looking at the direction of inconsistency relative to the SpeX data. In other words, we look at whether the NEOWISE data are inconsistent with the models that fit the SpeX data because the data are too warm or too cool relative to the model. QE2 and FG3 tend to push the NEOWISE models toward higher visible albedos, meaning that the data are cooler than we would expect relative to the SpeX data. Conversely, SD220 tends to have W2 points that are too warm relative to the model. Similarly, the UO1 set that is inconsistent has albedos that are too low, implying that the data are warmer than expected. In cases where JD6 data are inconsistent or no fit, it is the W1 point that is too warm. These groupings (too cool vs. too warm) are similar to those we see when looking at the fit consistencies, with a clear demarcation between fainter and brighter objects, as measured by average NEOWISE magnitude.

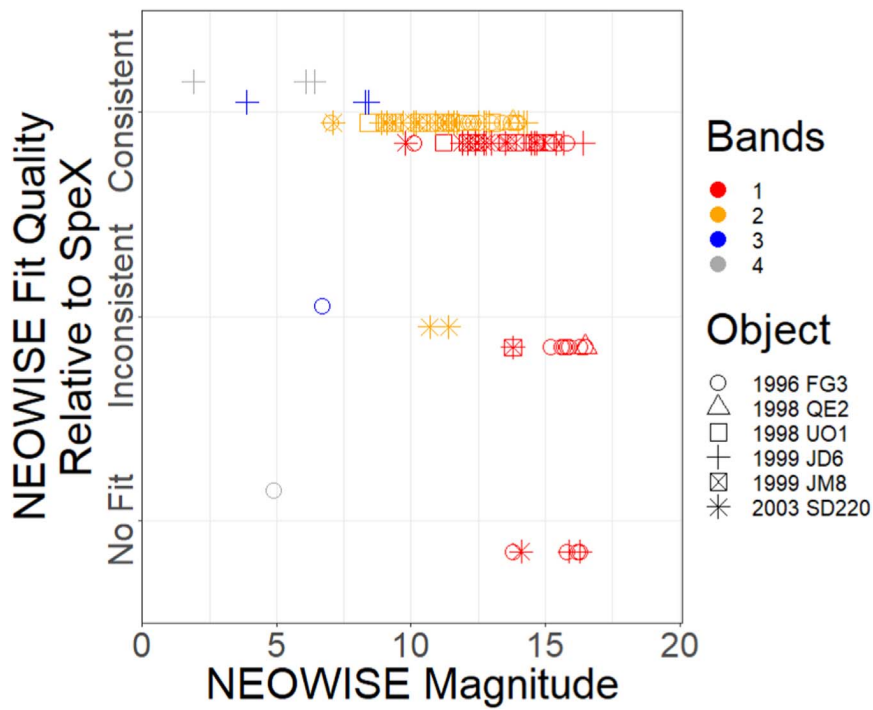

**Figure 16.** The consistency of NEOWISE model fits relative to models based on the SpeX data as a function of the NEOWISE magnitude. Each point is a different NEOWISE band observation (i.e., here the W1 and W2 band observations from a single data set are treated separately). "Consistent" means that the model results based on that band are consistent with results based on SpeX data. "Inconsistent" means that a model was fit to the band but the model was inconsistent with the results based on the SpeX data. "No Fit" means that no models could be fit to the band. A band observation is characterized as consistent vs. inconsistent or inconsistent vs. no fit if at least one band combination involving the given band can be fit with a model that meets the given criteria. The different colors represent the different bands. Magnitude uncertainties are the same size as or smaller than the symbols. Bands have been offset for clarity.

We examine similar trends for the beaming parameter. Here we note that SD220 has NEOWISE data sets that, contrary to the above, push toward higher beaming parameters, thus indicating data that are cooler than we would expect. We note, though, that SD220 is an object for which we expect more extreme beaming parameter values, due to its highly elongated (and thus nonspherical) shape, as well as the high phase angles at which it was observed. Therefore, we expect shape effects and self-shadowing to play a major role in affecting its thermal properties, neither of which are modeled directly by our NEATM-like model. These shape and self-shadowing effect can therefore only be captured by adjustments in the beaming parameter. This is confirmed by the fact that two additional data sets can be found to fit when looking at models generated assuming a different $H$ magnitude associated with a smaller diameter.

### 5.2. Fit Consistency by NEOWISE Band

We also examine the fit consistency as a function of the NEOWISE bands analyzed. The goal here is to identify whether any bands are more or less likely to have inconsistencies with the SpeX data. Figure 16 plots fit consistency relative to the SpeX data as a function of NEOWISE magnitude, with the different bands highlighted. First, we note a potential correlation with fit consistency and NEOWISE magnitude, but with the existence of a few outlier points. The outlier points characterized as no fit and inconsistent but with very low (bright) magnitudes are the single FG3 WISE data set. There is no immediate reason why we expect this data set to behave differently, other than to note that the uncertainty on the W4 point especially is quite low (Figure 4), and there may be additional uncertainties that we are not accounting for. However, a trend is still seen such that it is only high (faint) magnitude bands for which a model cannot be fit to the data. This is in line with our previous discussion where fainter objects were identified as more likely to produce no fit or inconsistent fits. All observations here have been screened by consultation with the MPC, the NEOWISE data processing pipeline, and manually checked to ensure that they only include real detections. Thus, we believe this trend to be real and not a data artifact.

A trend in fit consistency relative to the SpeX data is also seen when looking at each band individually. Namely, we see that the W2 band is much more likely to result in consistent fits than the W1 band. This may be because the W1 band is more contaminated with reflected light than the W2 band. However, if this were the case, we would expect the lower-albedo objects, which have less reflected light contaminating the W1 band, to be more consistent, contrary to the results seen here. Therefore, another explanation is required. We note that a very loose but similar trend may be present for bands W3 and W4. However, identifying a trend with these bands is more difficult, as there are far fewer W3 and W4 data points examined here than W1 and W2 data points.

In general, we also observe that different bands tend to produce fits more consistent with the SpeX data than others, especially for the more problematic objects. For example, Figures 14 and 15 show that for FG3 and QE2 fits to the W2 band alone are more consistent with the SpeX data than fits to the W1 and W2 bands together. It is, of course, expected that fitting one versus two data points would produce a larger fit space; however, it is notable that the fit space is shifted in the direction of consistency.

### 5.3. Related Findings and Implications

The above analysis shows that some simple thermal model results based on NEOWISE data may not be consistent with those found with SpeX data. While the majority of data sets examined here are consistent, roughly 40% were not. This is true even when the SpeX results are consistent with other methods of analysis, including other simple and more complex thermal models, as well as sizes from radar observations. These results fit in with a larger picture showing potential issues with asteroid properties derived using NEOWISE data for individual objects. Many studies have previously found discrepancies between asteroid diameters derived using NEOWISE data and diameters derived using other methods (E. S. Howell & M. Nolan et al. 2012; P. Taylor et al. 2014; J. R. Masiero et al. 2019; P. Taylor et al. 2019; J. R. Masiero et al. 2021; S. A. Myers et al. 2023). Furthermore, there is work that suggests that significant (>10%) differences in diameter can be found when using different clusters of NEOWISE bands compared to using all four bands together (E. A. Whittaker et al. 2023). Specifically, E. A. Whittaker et al. (2023) find that using bands W2 and W3 gives smaller diameters and using bands W3 and W4 gives larger diameters. This is similar to the trend we identify here, that bands W3 and W4 may be more consistent with models based on SpeX data and that W1 in particular is less consistent.

Limitations in the NEOWISE data based on the solar distance of the objects observed have also been previously noted. J. M. Bauer et al. (2013) analyzed observations of 52 different centaurs with a population mean solar distance of 10.5 au using WISE. First, only five of these observations had any usable detections in the W1 and W2 bands after stacking images. Second, they performed a calculation of the expected signal-to-noise ratio in the different bands for a low-albedo object. They report that the signal is expected to be greatest in the W2 band out to ∼1 au, greatest in the W3 band up to ∼2.3 au, and greatest in the W4 band beyond 2.3 au. We have no observations in our data set beyond 2.0 au, and therefore we cannot directly test these results. However, we do notice a marked decrease in the consistency of the W1 band. Furthermore, the majority of our observations are clustered between 1 and 1.5 au, around where the W2 point is expected to have the highest signal-to-noise ratio, and we indeed find that the W2 point produces the most consistent fits for many of our objects.

Previous work has also found discrepancies between modeling based on NEOWISE data and modeling based on other wavelength regimes. The existence of a flux deficit between models based on measurements in the infrared region versus those made in the millimeter or submillimeter region has long been recognized (K. Johnston et al. 1982; W. Webster et al. 1988). More recently, however, J. Orlowski-Scherer et al. (2024) specifically find a flux deficit between thermophysical models based on WISE data and thermophysical models based on millimeter observations at Atacama. This study is done on a population of ∼200 objects. The deficits they find range between ∼5% and ∼30% depending on the wavelengths investigated. Furthermore, they find that the deficit is strongest for S-type objects and weakest for X-type objects (as defined by D. J. Tholen & M. A. Barucci 1989), which in fact show a flux excess relative to the WISE data. While we find that our lower-albedo objects also tend to be too warm, we notice a similar trend with our higher-albedo objects as well. This is particularly notable for SD220.

Discrepancies have also been found between thermal modeling done based on WISE data and that based on other IR instruments. T. G. Müller et al. (2023) examined the main belt asteroid (10920) 1998 BC1 using both NEATM and a more complex thermophysical model looking at WISE data and JWST MIRI data. Previous results based on NEATM modeling with WISE data had shown an albedo of 0.08, beaming parameter of 1.09, and spherical equivalent diameter of 15.7 km (A. Mainzer et al. 2019). However, T. G. Müller et al. (2023) found that this model was inconsistent with the MIRI data, which instead required an albedo of 0.08, beaming parameter of 0.76, and spherical equivalent diameter of 10 km to fit (see their Figure 14). This discrepancy again shows the potential issues with relying on simple thermal model results based on NEOWISE data for individual objects. It is worth noting that T. G. Müller et al. (2023) find a spherical equivalent diameter of 15.3 km and albedo of 0.05–0.09 when using a more complex thermophysical model that accounted for the object's shape. However, the initial simple thermal model fit with the WISE data could not be reconciled with the MIRI data without use of the more complex model. We also note that 1998 BC1, as a lower-albedo object, is the type of object for which we might expect more discrepancies based on our results.

A full accounting of why simple thermal models may be producing the potential discrepancies described here will require detailed modeling of many more objects using many different data sets with a more complex thermophysical model. As such, this accounting is outside the scope of this work. However, we can hypothesize a few potential mechanisms that may be at work. The key assumptions made by our NEATM-like model include assuming a spherical shape for the object and no surface roughness. These assumptions could be particularly problematic for the NEA population, which includes many binaries and whose members are often observed at high phase angles. Thus, objects with elongated shapes, such as contact binaries, will often have surfaces largely influenced by self-shadowing effects. Similarly, separated binaries may have instances of shadowing of the primary by the secondary (although this specific causal mechanism has been ruled out for the objects presented here). Furthermore, surface roughness is a key parameter that is completely ignored by this model, except as it affects phase angle corrections. Our model also assumes prograde rotation, and some of the objects examined here are known to be retrograde rotators (although this would not produce a strong effect for low thermal inertia objects). It is therefore likely that one or many of these effects are responsible for the discrepancies shown here.

Overall, we emphasize that, despite these findings, there are many objects for which NEOWISE-derived parameters have been found to be consistent with parameters derived via other methods. We also reemphasize that 60% of the data sets examined here were found to be consistent with other observations. However, this work shows that, with even a small sampling of specific NEAs, some clear discrepancies that could lead to mischaracterization can be found. Even if these discrepancies occur in only a small subset of objects, they have the potential to affect our understanding of some NEAs. For instance, if low-albedo objects are characterized as having higher visible albedos, this would underestimate their size and thus their potential threat to Earth. These potential mischaracterizations may also occur more frequently in the future as new facilities such as NEO Surveyor (A. Mainzer et al. 2022) and the Vera Rubin Observatory (R. L. Jones et al. 2015) come online, producing large quantities of NEA detections that it will only be feasible to analyze using simple thermal models like ours.

## 6. Summary

The purpose of this work is to explore potential inconsistencies between simple thermal model results based on SpeX data and model results based on NEOWISE data. This was motivated by previous studies that have found such inconsistencies between NEOWISE data and other data sets. To explore these inconsistencies, we modeled a variety of NEAs using a simple thermal model based on the standard NEATM model (A. W. Harris 1998; E. Howell & C. Magri et al. 2018; S. A. Myers et al. 2023). The model we use allows us to explore three different free parameters: the visible albedo, the thermal inertia, and the beaming parameter, a nonphysical scaling factor that provides a first-order accounting for the model simplifications.

Using this model, we model data of six different NEAs that represent a range of asteroid characteristics: (53319) 1999 JM8, (85989) 1999 JD6, (137032) 1998 UO1, (163899) 2003 SD220, (175706) 1996 FG3, and (285263) 1998 QE2. Together, this sample includes S- and C-complex objects, binary and singular objects, spherical and highly elongated objects, and fast and slow rotators. For every object, data were collected using both the SpeX instrument on the NASA IRTF (J. T. Rayner et al. 2003) and the NEOWISE spacecraft (A. Mainzer et al. 2011, 2014a). NEATM-like models were then fit to each data set, and models were compared to each other.

Our key findings are as follows:

1. *Simple thermal models based on IRTF SpeX data may be inconsistent with simple thermal models based on NEOWISE data for some objects.* In general, these inconsistencies are more pronounced for fainter objects than brighter ones. That is, fainter objects, especially objects with lower albedos, are more likely to have inconsistent model results. Modeled albedos for these objects were between 0.02 and 0.13 higher for NEOWISE data compared to SpeX data. This result was seen across two key characteristics:
   (a) *Composition.* More primitive objects with lower visible albedos were more likely to have inconsistent fits, or were unable to be fit at all. Of the six objects examined here JM8, FG3, and QE2 have more primitive compositions. FG3 and QE2 were the objects that most strongly gave inconsistent or no fit results.
   (b) *Magnitude.* NEOWISE observations with higher (fainter) measured NEOWISE apparent magnitudes were more likely to have inconsistent fits, or were unable to be fit at all ($\approx$63% of W1 observations fainter than magnitude 15 vs. $\approx$14% of W1 observations brighter than magnitude 15). Although strongest for W1, this was true both across all observations and within specific bands.
2. *These results are part of a larger set of studies that show potential inconsistencies between models based on NEOWISE data and models based on other data sets.* Together, they highlight a need to better understand the NEOWISE data set and its limitations. Although the NEOWISE data produce accurate results both on the population average and for the majority of individual objects, our work shows that results for some individual objects may be skewed. Furthermore, we note that there are many instances where the NEOWISE image processing pipeline did not flag spurious or contaminated images. Out of our 68 original NEOWISE data sets, roughly 13% were rejected for being made up entirely of spurious or contaminated images, and an additional 51% of the averaged magnitudes were adjusted by removing problematic images with spurious or contaminated data.
3. *These results have potential implications for our understanding of small-body populations.* We find that fainter objects may have overly inflated inferred visible albedos and thus reduced diameter measurements, potentially underestimating their threat level. Potential systematic errors in the inferred properties of an object are also important for applications that consider asteroids as individual bodies, such as mission planning or resource utilization.
4. *These results are important for future observations.* New asteroid surveys, such as NEO Surveyor and the Vera Rubin Observatory, are expected to drastically increase the number of known NEAs (R. L. Jones et al. 2015; A. Mainzer et al. 2022). Detailed analysis and follow-up of all of these objects will be impossible; therefore, our understanding of these objects will rely heavily on the use of simple thermal models as applied to data like those taken by NEOWISE. Understanding any potential issues and limitations with these models and data sets is therefore critical for obtaining a reliable understanding of small-body populations in the future.

Moving forward, work should focus on further validating and understanding these potential inconsistencies, as well as probing the underlying mechanisms behind them. This will require extensive modeling of many NEAs and comparing results with ground-based data and NEOWISE data. Additionally, modeling using both simple thermal models and more complex thermophysical models will be required to understand the exact model assumptions and limitations that may be driving these inconsistencies. Carrying out this work will be crucial to ensure that we can make full use of existing and future NEA observations. Without fully understanding the limitations inherent in the models and data sets being used to understand NEAs, we will not be able to fully utilize them to answer key questions about planet formation, volatile delivery to Earth, orbital evolution, and planetary defense.

## Acknowledgments

This work was partially funded by the NASA YORPD program (NASA grant 80NSSC21K0658). This material is based upon work supported by the National Science Foundation Graduate Research Fellowship Program under grant No. DGE-2137419. Any opinions, findings, and conclusions or recommendations expressed in this material are those of the author(s) and do not necessarily reflect the views of the National Science Foundation. Authors S.A.M., E.S.H., R.J.V. Jr., Y.R.F., and M.L.H. were Visiting Astronomers at the Infrared Telescope Facility, which is operated by the University of Hawaii under contract 80HQTR19D0030 with the National Aeronautics and Space Administration. S.E.M. was funded by NASA grant 80NSSC19K0523. Thanks to Jenna L. Crowell, Mariah Law, and Kiana McFadden for additional IRTF observing assistance. Additional thanks to the anonymous reviewers, whose comments greatly improved this manuscript.

*Software:* ggplot2 (H. Wickham 2016), dplyr (H. Wickham et al. 2023), gWidgets2 (J. Verzani 2013), Spextool (M. Cushing et al. 2004).

## ORCID iDs

Samuel A. Myers https://orcid.org/0000-0001-8500-6601
Ellen S. Howell https://orcid.org/0000-0002-7683-5843
Christopher Magri https://orcid.org/0000-0002-2200-4622
Ronald J. Vervack, Jr. https://orcid.org/0000-0002-8227-9564
Yanga R. Fernández https://orcid.org/0000-0003-1156-9721
Mary L. Hinkle https://orcid.org/0000-0003-0713-2252
Sean E. Marshall https://orcid.org/0000-0002-8144-7570

## References

Azadmanesh, M., Roshanian, J., & Hassanalian, M. 2023, On the Importance of Studying Asteroids: A Comprehensive Review, PrAeS, 142, 100957
Bauer, J. M., Grav, T., Blauvelt, E., et al. 2013, Centaurs and Scattered Disk Objects in the Thermal Infrared: Analysis of WISE/NEOWISE Observations, ApJ, 773, 22
Benner, L. A., Brozovic, M., Giorgini, J., et al. 2012, AAS/DPS Meeting Abstracts, 44, 102.06
Benner, L. A., Busch, M., Giorgini, J., Taylor, P., & Margot, J.-L. 2015, Radar Observations of Near-Earth and Main-Belt Asteroids, in Asteroids IV, ed. P. Michel, F. E. DeMeo, & W. F. Bottke (Tucson, AZ: Univ. Arizona Press), 165
Benner, L. A., Ostro, S. J., Nolan, M. C., et al. 2002, Radar Observations of Asteroid 1999 JM8, M&PS, 37, 779
Binzel, R. P., DeMeo, F. E., Turtelboom, E. V., et al. 2019, Compositional Distributions and Evolutionary Processes for the Near-Earth Object Population: Results from the MIT-Hawaii Near-Earth Object Spectroscopic Survey (MITHNEOS), Icar, 324, 41
Binzel, R. P., Harris, A. W., Bus, S. J., & Burbine, T. H. 2001, Spectral Properties of Near-Earth Objects: Palomar and IRTF Results for 48 Objects Including Spacecraft Targets (9969) Braille and (10302) 1989 ML, Icar, 151, 139
Binzel, R. P., Reddy, V., & Dunn, T. 2015, The Near-Earth Object Population: Connections to Comets, Main-belt Asteroids, and Meteorites, in Asteroids IV, ed. P. Michel, F. E. DeMeo, & W. F. Bottke (Tucson, AZ: Univ. Arizona Press), 243
Bondarenko, Y. S., Marshalov, D., Vavilov, D., & Medvedev, Y. D. 2019, Radar observations of near-Earth asteroid 2003 SD220, EPSC-DPS Joint Meeting, 13, 554
Bottke, W. F., Jr., Cellino, A., Paolicchi, P., & Binzel, R. P. 2002, An Overview of the Asteroids: The Asteroids III Perspective, in Asteroids III, ed. W. F. Bottke, Jr. et al. (Tucson, AZ: Univ. Arizona Press), 3
Bowell, E., Hapke, B., Domingue, D., et al. 1989, Application of Photometric Models to Asteroids, in Asteroids II, ed. R. P. Binzel, T. Gehrels, & M. S. Matthews (Tucson, AZ: Univ. Arizona Press), 524
Campins, H., Kelley, M. S., Fernandez, Y., Licandro, J., & Hargrove, K. 2009, Low Perihelion Near-Earth Asteroids, EM&P, 105, 159
Cushing, M., Vacca, W., & Rayner, J. 2004, Spextool: A Spectral Extraction Package for SpeX, a 0.8–5.5 Micron Cross-Dispersed Spectrograph, PASP, 116, 362
Cutri, R., Mainzer, A., Conrow, T., et al. 2015, Explanatory Supplement to the NEOWISE Data Release Products, IPAC, https://wise2.ipac.caltech.edu/docs/release/neowise/expsup/
Delbo, M., Mueller, M., Emery, J. P., Rozitis, B., & Capria, M. T. 2015, Asteroid Thermophysical Modeling, in Asteroids IV, ed. P. Michel, F. E. DeMeo, & W. F. Bottke (Tucson, AZ: Univ. Arizona Press), 107
Fieber-Beyer, S. K., Kareta, T., Reddy, V., & Gaffey, M. J. 2020, Near-Earth Asteroid:(285263) 1998 QE2, Icar, 347, 113807
Grav, T., Mainzer, A. K., Masiero, J. R., et al. 2023, The NEO Surveyor Near-Earth Asteroid Known Object Model, PSJ, 4, 228
Harris, A. W. 1998, A Thermal Model for Near-Earth Asteroids, Icar, 131, 291
Harris, A. W., Boslough, M., Chapman, C. R., Drube, L., & Michel, P. 2015, Asteroid Impacts and Modern Civilization: Can We Prevent a Catastrophe?, in Asteroids IV, ed. P. Michel, F. E. DeMeo, & W. F. Bottke (Tucson, AZ: Univ. Arizona Press), 835
Hicks, M., Lawrence, K., Chesley, S., et al. 2013, Palomar Spectroscopy of Near-Earth Asteroids 137199 (1999 KX4), 152756 (1999 JV3), 163249 (2002 GT), 163364 (2002 OD20), and 285263 (1998 QE2)., ATel, 5132, 1
Horiuchi, S., Molyneux, B., Stevens, J. B., et al. 2021, Bistatic Radar Observations of Near-Earth Asteroid (163899) 2003 SD220 from the Southern Hemisphere, Icar, 357, 114250
Howell, E., Magri, C., Vervack, R., Jr, et al. 2018, SHERMAN–A Shape-based Thermophysical Model II. Application to 8567 (1996 HW1), Icar, 303, 220
Howell, E. S., Magri, C., Vervack, R., et al. 2008, Thermal Infrared Observations of Several Near-Earth Asteroids, AAS/DPS Meeting, 40, 28.02
Howell, E. S., Vervack, R., Jr, Nolan, M., et al. 2012, Combining Thermal and Radar Observations of Near-Earth Asteroids, AAS/DPS Meeting, 44, 110.07
Johnston, K., Seidelmann, P., & Wade, C. 1982, Observations of 1 Ceres and 2 Pallas at Centimeter Wavelengths, AJ, 87, 1593
Jones, R. L., Juric, M., & Ivezic, Ž. 2015, Asteroid Discovery and Characterization with the Large Synoptic Survey Telescope, in IAU Symp. 318, Asteroids: New Observations, New Models (Cambridge: Cambridge Univ. Press), 282
Juric, M., Ivezic, Ž., Lupton, R. H., et al. 2002, Comparison of Positions and Magnitudes of Asteroids Observed in the Sloan Digital Sky Survey with Those Predicted for Known Asteroids, AJ, 124, 1776
Kuroda, D., Ishiguro, M., Naito, H., et al. 2021, (85989) 1999 JD6: a first Barbarian asteroid detected by polarimetry in the NEA population, A&A, 646, A51
Lebofsky, L. A., & Spencer, J. R. 1989, Radiometry and Thermal Modeling of Asteroids, in Asteroids II, ed. R. P. Binzel, T. Gehrels, & M. S. Matthews (Tucson, AZ: Univ. Arizona Press), 128
Lebofsky, L. A., Sykes, M. V., Tedesco, E. F., et al. 1986, A Refined "Standard" Thermal Model for Asteroids Based on Observations of 1 Ceres and 2 Pallas, Icar, 68, 239
Lewin, C. D., Howell, E. S., Vervack, R. J., et al. 2020, Near-infrared Spectral Characterization of Solar-type Stars in the Northern Hemisphere, AJ, 160, 130
Lord, S. D. 1992, Computing Earth's Atmospheric Transmission of Near-and Far-infrared Radiation NASA-TM-103957, NASA Ames Research Ctr. https://ntrs.nasa.gov/citations/19930010877
Magri, C., Howell, E. S., Vervack, R. J., Jr, et al. 2018, SHERMAN, A Shape-based Thermophysical Model I. Model Description and Validation, Icar, 303, 203
Mainzer, A., Bauer, J., Cutri, R., et al. 2014a, Initial Performance of the NEOWISE Reactivation Mission, ApJ, 792, 30
Mainzer, A., Bauer, J., Cutri, R., et al. 2019, NEOWISE Diameters and Albedos V2.0, NASA Planetary Data System, urn:nasa:pds:neowise_diameters_albedos::2.0, doi:10.26033/18S3-2Z54
Mainzer, A., Bauer, J., Grav, T., et al. 2011, Preliminary Results from NEOWISE: An Enhancement to the Wide-field Infrared Survey Explorer for Solar System Science, ApJ, 731, 53
Mainzer, A., Bauer, J., Grav, T., et al. 2014b, The Population of Tiny Near-Earth Objects Observed by NEOWISE, ApJ, 784, 110
Mainzer, A., Masiero, J., Wright, E., et al. 2022, The Near-Earth Object Surveyor Mission, AAS/DPS Meeting, 54, 409.02
Mainzer, A. K., Masiero, J. R., Abell, P. A., et al. 2023, The Near-Earth Object Surveyor Mission, PSJ, 4, 224
Marshall, S., Howell, E., Nolan, M., et al. 2014, Near-Earth asteroid (137032) 1998 UO_1: Shape Model and Thermal Properties, in Asteroids, Comets, Meteors Conf., 337

Marshall, S. E., Howell, E. S., Brozović, M., et al. 2015, Potentially Hazardous Asteroid (85989) 1999 JD6: Radar, Infrared, and Lightcurve Observations and a Preliminary Shape Model, AAS/DPS Meeting, 47, 204.09

Marshall, S. E., Howell, E. S., Vervack, R. J., Jr, et al. 2017, Thermophysical Modeling of Potentially Hazardous Asteroid (85989) 1999 JD6, AAS/DPS Meeting, 49, 110.15

Masiero, J. R., Mainzer, A. K., Bauer, J. M., et al. 2020, Asteroid Diameters and Albedos from NEOWISE Reactivation Mission Years 4 and 5, PSJ, 1, 5

Masiero, J. R., Mainzer, A. K., Bauer, J. M., et al. 2021, Asteroid Diameters and Albedos from NEOWISE Reactivation Mission Years Six and Seven, PSJ, 2, 162

Masiero, J. R., Wright, E. L., & Mainzer, A. K. 2019, Thermophysical Modeling of NEOWISE Observations of DESTINY+ Targets Phaethon and 2005 UD, AJ, 158, 97

Michel, P., DeMeo, F. E., & Bottke, W. F. 2015, Asteroids IV (Tucson, AZ: Univ. Arizona Press)

Moeyens, J., Myhrvold, N., & Ivezic, Ž. 2020, ATM: An Open-source Tool for Asteroid Thermal Modeling and Its Application to NEOWISE Data, Icar, 341, 113575

Moskovitz, N. A., Polishook, D., DeMeo, F. E., et al. 2017, Near-infrared Thermal Emission from Near-Earth Asteroids: Aspect-dependent Variability, Icar, 284, 97

Mueller, M., Delbo, M., Hora, J., et al. 2011, ExploreNEOs. III. Physical Characterization of 65 Potential Spacecraft Target Asteroids, AJ, 141, 109

Müller, T. G., Micheli, M., Santana-Ros, T., et al. 2023, Asteroids Seen by JWST-MIRI: Radiometric Size, Distance, and Orbit Constraints, A&A, 670, A53

Myers, S. A., Howell, E. S., Magri, C., et al. 2023, Constraining the Limitations of NEATM-like Models: A Case Study with Near-Earth Asteroid (285263) 1998 QE2, PSJ, 4, 5

NEOWISE Team 2020, NEOWISE-R Single Exposure (L1b) Source Table, NASA IPAC DataSet, IRSA144 doi:10.26131/IRSA144

Nolan, M. C., Magri, C., Howell, E. S., et al. 2013, Shape Model and Surface Properties of the OSIRIS-REx Target Asteroid (101955) Bennu from Radar and Lightcurve Observations, Icar, 226, 629

Nugent, C. R., Mainzer, A., Bauer, J., et al. 2016, NEOWISE Reactivation Mission Year Two: Asteroid Diameters and Albedos, AJ, 152, 63

Orlowski-Scherer, J., Venterea, R., Battaglia, N., et al. 2024, The Atacama Cosmology Telescope: Millimeter Observations of a Population of Asteroids or: ACTeroids, ApJ, 964, 138

Ostro, S. J. 1985, Radar Observations of Asteroids and Comets, PASP, 97, 877

Polishook, D., & Brosch, N. 2008, Photometry of Aten asteroids–More than a Handful of Binaries, Icar, 194, 111

Pravec, P., Šarounová, L., Rabinowitz, D. L., et al. 2000, Two-period Lightcurves of 1996 FG3, 1998 PG, and (5407) 1992 AX: One Probable and two Possible Binary Asteroids, Icar, 146, 190

Rayner, J. T., Toomey, D. W., Onaka, P. M., et al. 2003, SpeX: a Medium-resolution 0.8–5.5 Micron Spectrograph and Imager for the NASA Infrared Telescope Facility, PASP, 115, 362

Reddy, V., Gaffey, M. J., Abell, P. A., & Hardersen, P. S. 2012, Constraining Albedo, Diameter and Composition of Near-Earth Asteroids via Near-infrared Spectroscopy, Icar, 219, 382

Rivera-Valentín, E., Taylor, P., Reddy, V., et al. 2019, Radar and Near-Infrared Characterization of Near-Earth Asteroid (163899) 2003 SD220, LPSC, 50, 3016

Scheirich, P., & Pravec, P. 2009, Modeling of Lightcurves of Binary Asteroids, Icar, 200, 531

Spencer, J. R., Lebofsky, L. A., & Sykes, M. V. 1989, Systematic Biases in Radiometric Diameter Determinations, Icar, 78, 337

Springmann, A., Taylor, P., Howell, E., et al. 2014, Radar Shape Model of Binary Near-Earth Asteroid (285263) 1998 QE2, LPSC, 45, 1313

Taylor, P., Howell, E., Nolan, M., et al. 2014, Comparing the Diameters and Visual Albedos Derived from Radar and Infrared Observations, in Asteroids, Comets, Meteors Conf., 524

Taylor, P., Rivera-Valentín, E., & Aponte-Hernandez, B. 2019, Shape Model of 3200 Phaethon from Radar and Lightcurve Observations, EPSC-DPS Joint Meeting, 13, 1246

Tholen, D. J., & Barucci, M. A. 1989, Asteroid Taxonomy, in Asteroids II, ed. R. P. Binzel, T. Gehrels, & M. S. Matthews (Tucson, AZ: Univ. Arizona Press), 298

Thomas, C. A., Trilling, D. E., Emery, J. P., et al. 2011, ExploreNEOs. V. Average Albedo by Taxonomic Complex in the Near-Earth Asteroid Population, AJ, 142, 85

Tian, J., Zhao, H.-B., & Li, B. 2022, Shape Model and Rotation Acceleration of (1685) Toro and (85989) 1999 JD6 from Optical Observations, RAA, 22, 125004

Trilling, D. E., Mueller, M., Hora, J. L., et al. 2010, ExploreNEOs. I. Description and First Results from the Warm Spitzer Near-earth Object Survey, AJ, 140, 770

Veres, P., Jedicke, R., Fitzsimmons, A., et al. 2015, Absolute Magnitudes and Slope Parameters for 250,000 Asteroids Observed by Pan-STARRS PS1-Preliminary Results, Icar, 261, 34

Verzani, J., 2013 gWidgets2: Rewrite of gWidgets API for Simplified GUI Construction, The R Foundation

Walsh, K. J., Mueller, M., Binzel, R. P., & DeMeo, F. E. 2012, Physical Characterization and Origin of Binary Near-Earth Asteroid (175706) 1996 FG3, ApJ, 748, 104

Warner, B. D. 2016, Near-Earth Asteroid Lightcurve Analysis At Cs3-palmer Divide Station: 2015 October-December, MPBu, 43, 143

Webster, W., Johnston, K., Hobbs, R., et al. 1988, The Microwave Spectrum of Asteroid Ceres, AJ, 95, 1263

Whittaker, E. A., Margot, J.-L., Lam, A. L. H., & Myhrvold, N. 2023, Thermal Models of Asteroids with Two-band Combinations of Wide-field Infrared Survey Explorer Cryogenic Data, PSJ, 4, 64

Wickham, H. 2016, ggplot2: Elegant Graphics for Data Analysis (New York: Springer)

Wickham, H., François, R., Henry, L., Müller, K., & Vaughan, D. 2023, dplyr: A Grammar of Data Manipulation, https://dplyr.tidyverse.org

WISE Team 2020a, NEOWISE 2-Band Post-Cryo Single Exposure (L1b) Source Table, NASA IPAC DataSet, IRSA124, doi:10.26131/IRSA124

WISE Team 2020b, WISE All-Sky Single Exposure (L1b) Source Table, NASA IPAC DataSet, IRSA139, doi:10.26131/IRSA139

Wolters, S. D., Green, S. F., McBride, N., & Davies, J. K. 2008, Thermal Infrared and Optical Observations of Four Near-Earth Asteroids, Icar, 193, 535

Wolters, S. D., Rozitis, B., Duddy, S. R., et al. 2011, Physical Characterization of Low Delta-V Asteroid (175706) 1996 FG3, MNRAS, 418, 1246

Wright, E. L., Eisenhardt, P. R. M., Mainzer, A. K., et al. 2010, The Wide-field Infrared Survey Explorer (WISE): Mission Description and Initial On-orbit Performance, AJ, 140, 1868

Yu, L., Ji, J., & Wang, S. 2014, Shape, Thermal and Surface Properties Determination of a Candidate Spacecraft Target Asteroid (175706) 1996 FG3, MNRAS, 439, 3357